\documentclass[12pt,pra,aps]{revtex4-1}

\usepackage{bm}
\usepackage{amssymb}
\usepackage{latexsym}
\usepackage{amsfonts}
\usepackage{amsmath}
\usepackage{epsfig}
\usepackage{psfrag}
\usepackage{xcolor}
\usepackage{hyperref}
\usepackage{natbib}
\newcommand{\ket}[1]{\left|#1\right>}
\newcommand{\bra}[1]{\left<#1\right|}

\newcommand{\expval}[1]{\left< #1 \right>}
\newcommand{\nn}{\nonumber\\}

\newcommand{\f}[1]{\mbox{\boldmath$#1$}}

\newcommand{\bea}{\begin{eqnarray}}
\newcommand{\eea}{\end{eqnarray}}
\newcommand{\ord}{{\cal O}}
\newcommand{\abs}[1]{{\left| #1 \right|}}
\newcommand{\trace}[1]{{\rm Tr}\left\{ #1 \right\}}
\newcommand{\traceR}[1]{{\rm Tr_R}\left\{ #1 \right\}}
\newcommand{\ii}{{\rm i}}

\allowdisplaybreaks[1]

\begin{document}

\title{Quantum Thermodynamics with Degenerate Eigenstate Coherences}

\author{G. Bulnes Cuetara$^1$}
\author{M. Esposito$^1$}
\author{G. Schaller$^2$}

\address{
$^{1}$  \quad Complex Systems and Statistical Mechanics, Physics and Materials Research Unit, University of Luxembourg, L-1511 Luxembourg, Luxembourg \\
$^{2}$ \quad Institut f\"ur Theoretische Physik, Technische Universit\"at Berlin, Hardenbergstr. 36, D-10623 Berlin, Germany
}

\begin{abstract}
We establish quantum thermodynamics for open quantum systems weakly coupled to their reservoirs when the system exhibits degeneracies.
The first and second law of thermodynamics are derived, as well as a finite-time fluctuation theorem for mechanical work and energy and matter currents.
Using a double quantum dot junction model, local eigenbasis coherences are shown to play a crucial role on thermodynamics and on the electron counting statistics. 
\end{abstract}

\maketitle

\section{Introduction}

The study of nonequilibrium open quantum systems is an active field of research with particular relevance to routinely devised systems such as quantum dots or electronic circuits~\cite{Fujisawa_2006_Science, gustavsson2009electron, kung2012irreversibility, saira2012test}, or assemblies of cold atoms~\cite{liebisch2005atom, malossi2014full, krinner2015observation}, for example. The possibility to monitor thermodynamically relevant quantities, such as heat and work, during a single experimental realization has motivated the study of their fluctuating properties and thereby, the identification of universal laws satisfied by their statistics~\cite{PekolaNatPhys15, Esposito_2009_ReviewsofModernPhysics, Campisi2011ReviewofModernPhysics}.

Quantum master equations have been widely used for the study of the thermodynamic properties of open quantum systems~\cite{SpohnLebowitz78, Esposito_2009_ReviewsofModernPhysics, Kosloff13, Kurizki15}. They are usually derived for systems weakly interacting with their reservoirs using the Born-Markov and secular (BMS) approximation~\cite{spohn1980kinetic, breuer2002theory, gardiner2004quantum}. The resulting quantum master equation can be shown to be of Lindblad form~\cite{lindblad1976generators}. In absence of degeneracies in the system Hamiltonian, the density matrix populations in the system energy eigenbasis satisfy a closed stochastic equation whereas coherences undergo an independent decay in time \cite{HarbolaEsposito06}. For many processes which only depend on populations, or for steady state dynamics where eigenstates coherences are always vanishing, a classical Stochastic Thermodynamics (ST)~\cite{Seifert_2012_ReportsonProgressinPhysics, van2015ensemble, esposito2012stochastic} can be easily build for the population dynamics. This provides a consistent framework for the study of the thermodynamics of open quantum systems at both the average and the single trajectory level \cite{Esposito_2007_PhysRevE, HarbolaEspositoBoson07, EspositoCuetaraGaspard11, AlonsoCorreaPRE13, EspoKrauseJCP15, EspoBulNJP15,  KosloffJCP84}. However, various time dependent processes do depend on eigenstate coherences. This happens for instance for systems driven by fast periodic time-dependent forces, where what we just said holds at the level of quasi-energies instead of eigenenergies \cite{EspoBulnEng15, Kurizki15}. It also happens for multi-stroke machines or for systems undergoing feedback control, where eigenstate coherences can be shown to play an important thermodynamics role 
(see e.g., \cite{Kosloff15PRX}, respectively \cite{Strasberg2013}). 

Open quantum systems with degenerate system energies constitute another important case in which eigenstate coherences come into play already in the weak coupling limit. In this case time-dependent driving is not even required and coherences may survive even at steady state. Such situations are very important in mesoscopic physics. 
The aim of this paper is to extend central results of stochastic thermodynamics to open quantum systems with degeneracies. 
When applying the BMS approximations, while the dynamics of populations and coherences between non-degenerate states of the system Hamiltonian remains uncoupled, 
the populations and coherences between degenerate states remain coupled.
We propose consistent definitions for energy, work, heat, entropy, and entropy production for such dynamics. 
We further obtain the counting statistics of the mechanical work and energy and particle currents from the aforementioned quantum master equation 
and derive a finite time fluctuation theorem which extends its classical counterpart~\cite{cuetara2014exact} to quantum systems with eigenstate coherences.
We illustrate our results on a degenerate double quantum dot system which exhibits a quantum suppression of the particle current due to coherences \cite{braun2004a, darau2009a,schultz2009a,schaller2009b,schultz2010a, nilsson2010a, karlstroem2011a}. We show that coherences cause a bi-modality in the finite time current distribution \cite{schaller2009b}, which is nevertheless compatible with the fluctuation theorem symmetry.


The paper is organized as follows. 
The BMS master equation for a general open quantum system with exact degeneracies 
is exposed in section~\ref{microDerivation}. The analysis of the nonequilibrium thermodynamics is presented in section~\ref{average_thermodynamics}, where we establish the energy and entropy balance, as well as the positivity of entropy production. The thermodynamics analysis is exposed in section \ref{stochastic_thermodynamics}. An expression for the work and currents statistics is derived in section \ref{counting_statistics} using the dressed quantum master equation formalism~\cite{Esposito_2009_ReviewsofModernPhysics}.
We prove a finite-time fluctuation theorem for systems described by the quantum master equation~(\ref{EQ:bms_lindblad}) in section~\ref{finiteTimeFT}. 
Finally, our approach is applied in section~\ref{a_quantum_master_equation_model} to study the thermodynamics of a degenerate double quantum dot connected to two electronic leads.
A summary is given in section~\ref{Conc}. 
%
\section{General Formalism}

\subsection{Microscopic derivation of Lindblad master equations}\label{microDerivation}

We consider an open quantum system with Hamiltonian $H=H_S+H_R+H_I$, in terms of system ($H_S$), 
reservoir ($H_R$), and interaction ($H_I$) Hamiltonians, respectively.
We aim to describe the
effective dynamics of the system with a master equation of the form
\bea \label{gen_master_eq}
\dot{\rho}_S = {\cal L} \rho_S 
\eea
for the system density matrix $\rho_S = \traceR{\rho}$ only (here and in the following, $\traceR{\ldots}$ denotes the partial
trace over the reservoir degrees of freedom).
This equation should preserve the density matrix properties (trace, hermiticity and positivity)
at least in an approximate sense.
The Lindblad master equation~\cite{lindblad1976generators} is the most general master equation that preserves the
density matrix properties exactly.
There are multiple ways of obtaining Lindblad master equations from microscopic Hamiltonians for various parameter regimes~\cite{lidar2001a,schaller2008a}.
Here, we will constrain ourselves to the weak-coupling limit between system and reservoir, in which the Born-, Markov-, and secular (BMS) approximations~\cite{breuer2002theory} 
can be applied, the latter often also termed rotating wave approximation. As such, we will be concerned with systems whose relaxation dynamics is much slower than the fast correlation time of the reservoirs. 

Under the aforementioned approximations, and for a decomposition of the interaction Hamiltonian 
\bea
H_I = \sum_\alpha A_\alpha \otimes B_\alpha
\eea
into system operators $A_\alpha$ and reservoir operators $B_\alpha$, respectively, 
the BMS Lindblad master equation becomes for a single reservoir~\cite{schaller2014}
\bea\label{EQ:bms_lindblad}
\dot\rho_S ( t) & = &
 -\ii \left[H_S + \sum_{ab} \sigma_{ab} L_{ab}, \rho_S ( t) \right]
+ \sum_{ab,cd} \gamma_{ab,cd} \left[ L_{ab} \rho_S ( t) L_{cd}^\dagger - \frac{1}{2} \left\{L_{cd}^\dagger L_{ab}, \rho_S ( t) \right\}\right] \\
& \equiv &  {\cal L} \rho_S ( t) \,, 
\eea
where we use the fixed eigen-operator basis $L_{ab} = \ket{a}\bra{b}$ of the system Hamiltonian $H_S \ket{a} = E_a \ket{a}$.
We note that this basis is unique when the spectrum of $H_S$ is non-degenerate.
Here, the matrix elements of the Lamb shift Hamiltonian $\sigma_{ab} = \sigma_{ba}^*$ and
the positive definite matrix $\gamma_{ab,cd}$ are given by
\bea\label{EQ:dampening_constants}
\sigma_{ab} &=& \delta_{E_b,E_a} \sum_{\alpha\beta} \sum_c \frac{\sigma_{\alpha\beta}(E_b-E_c)}{2\ii}  \bra{c} A_\beta \ket{b} \bra{c} A_\alpha^\dagger \ket{a}^*\,,\nn
\gamma_{ab,cd} &=& \delta_{E_b-E_a,E_d-E_c} \sum_{\alpha\beta} \gamma_{\alpha\beta}(E_b-E_a) \bra{a} A_\beta \ket{b} \bra{c} A_\alpha^\dagger \ket{d}^*\,.
\eea
They depend on the matrix elements of the system coupling operators $A_\alpha$ 
and the even ($\gamma_{\alpha\beta}$) and odd ($\sigma_{\alpha\beta}$) Fourier transforms 
\bea\label{EQ:fourier_transforms}
\gamma_{\alpha\beta}(\omega) = \int C_{\alpha\beta}(\tau) e^{+\ii\omega\tau} d\omega\,,\qquad
\sigma_{\alpha\beta}(\omega) = \int {\rm sgn}(\tau) C_{\alpha\beta}(\tau) e^{+\ii\omega\tau} d\omega
\eea
of the reservoir correlation functions (bold symbols denote the interaction picture $\f{B_\alpha}(\tau) = e^{+\ii H_R\tau} B_\alpha e^{-\ii H_R\tau}$)
\bea
C_{\alpha\beta}(\tau) = \expval{\f{B_\alpha}(\tau) B_\beta} = \traceR{\f{B_\alpha}(\tau) B_\beta \bar\rho_R}\,,
\eea
where $\bar\rho_R$ denotes the stationary state of the reservoir.
For a single reservoir it is usually chosen as a thermal reference state
\bea\label{EQ:gibbs_reservoir}
\bar\rho_R = e^{-\beta(H_R - \mu N_R - \phi_R)}
\eea
in terms of the reservoir thermodynamic grand-potential $\phi_R = - \beta^{-1} \ln \mbox{Tr} \left\{ e^{-\beta(H_R - \mu N_R )} \right\}$.
It is characterized by the inverse temperature $\beta$ and chemical potential $\mu$ of the reservoir.

We now summarize a few useful properties of the BMS Lindblad master equation beyond preservation of the density matrix properties.

First, we observe that coherences $\rho_{ij} \equiv \langle i | \rho_S (t) | j \rangle$ of basis states $i$ and $j$ with different energies $E_i \neq E_j$ will evolve decoupled from the
populations $\rho_{aa} \equiv \langle a | \rho_S (t) | a\rangle$
\bea
{\dot\rho}_{ij} = -\ii \left(E_i-E_j+\sigma_{ii}-\sigma_{jj}\right) \rho_{ij} 
+ \sum_{ab} \gamma_{ia,jb} \rho_{ab} 
- \frac{1}{2} \sum_{ab} \gamma_{ab,ai} \rho_{bj} - \frac{1}{2} \sum_{ab} \gamma_{aj,ab} \rho_{ib}\,,
\eea
which formally results from the Kronecker-delta functions in~(\ref{EQ:dampening_constants}).
This implies that for a non-degenerate system (where $E_i \neq E_j$ implies $i \neq j$), one can directly show that in the system energy
eigenbasis the master equation decouples the evolution of all populations and all coherences.
Whereas the coherences are damped and will fade away in the long-term limit, the equation governing the dynamics of populations
in this case just becomes a simple rate equation with transition rates from $b$ to $a$ given by
$\gamma_{ab,ab}$
\bea \label{rwa_mast_eq}
\dot \rho_{aa} = \sum_b \gamma_{ab,ab} \rho_{bb} - \sum_b \gamma_{ba,ba} \rho_{aa}\,.
\eea
Instead, for states with same energies the populations of the system density matrix are coupled to the coherences of the states with the same energy. 
The treatment which disregards all couplings of the populations to the coherences will in this paper be denoted 
the {\it rotating wave approximation} (RWA). 
In contrast, the treatment which preserves the couplings to the degenerate coherences will be denoted the {\it secular approximation}  (BMS).

Second, for a single reservoir in thermal equilibrium~(\ref{EQ:gibbs_reservoir}), the correlation functions acquire
additional analytic properties, so-called Kubo-Martin-Schwinger relations (KMS), which enable a thermodynamically consistent
description even for degenerate systems.
In absence of chemical potentials, the KMS relations read
\bea
C_{\alpha\beta}(\tau) = C_{\beta\alpha}(-\tau-\ii\beta)\,,
\eea
and transfer to the even Fourier transforms as $\gamma_{\alpha\beta}(+\omega) = \gamma_{\beta\alpha}(-\omega) e^{+\beta\omega}$.
In the master equation, these eventually lead to detailed balance, and the system thermalizes in the long run
with the temperature of the reservoir, i.e., $\bar\rho_S \propto e^{-\beta H_S}$ is
a stationary state of the master equation~\cite{breuer2002theory}.
With a chemical potential and an interaction that conserves the total particle number, i.e., under the assumption
that $[H_S, N_S] = [H_R, N_R] = [H_I, N_S + N_R]=0$, the KMS relations can be generalized to
\bea\label{EQ:kms_chempot}
\sum_{\bar\alpha} A_{\bar\alpha} C_{\alpha\bar\alpha}(\tau) = \sum_{\bar\alpha} e^{+\beta\mu N_S} A_{\bar\alpha} e^{-\beta\mu N_S} C_{\bar\alpha\alpha}(-\tau-\ii\beta)\,,
\eea
which is explicitly shown in Appendix~\ref{APP:kms_chempot}.
This leads to the local detailed detailed balance (LDB) relation among the coefficients
\bea\label{EQ:detailed_balance}
\frac{\gamma_{ab,cd}}{\gamma_{dc,ba}} = e^{\beta\left[(E_b-E_a)-\mu(N_b-N_a)\right]}\,,
\eea
where $E_a$ and $N_a$ denote energy and particle number of state $a$, respectively.
Eventually, these relations imply equilibration of both the temperature and chemical potential~\cite{schaller2011a}, i.e., 
$\bar\rho_S = e^{-\beta (H_S - \mu N_S - \phi_S )}$, where $\phi_S = -\beta^{-1} \ln \mbox{Tr} \left\{ \exp{- \beta (H_S - \mu N_S)} \right\}$, 
is one stationary state of the BMS master equation -- even in presence of degeneracies.

We extend our setup by admitting two kinds of drivings.

First, we may allow for a slow external driving of the system Hamiltonian $H_S \to H_S(t)$. 
However, this driving must be significantly slower than the decay time of the reservoir correlation functions. 
Furthermore, the driving should not lift the degeneracy in the energy spectrum, and the non-degenerate states should not cross at any time. 
In other words, the driving should only operate on well separated eigenenergies.  
Under these assumptions our approximations remains applicable, and we still arrive at the same microscopically derived master equation. 
The only difference is that the previously constant Hamiltonian and all associated quantities become time-dependent:
$H_S \to H_S(t)$, $E_a \to E_a(t)$, $L_{ab} \to L_{ab}(t)$, and $\ket{a} \to \ket{a(t)}$ in Eqns.~(\ref{EQ:bms_lindblad}) and~(\ref{EQ:dampening_constants}).

Second, we can consider $N$ multiple reservoirs $H_R = \sum_\nu H_R^{(\nu)}$ held at different equilibrium states
\bea\label{EQ:gibbs_reservoir_mult}
\bar\rho_R = \bigotimes_\nu e^{-\beta_\nu (H_R^{(\nu)} - \mu_\nu N_R^{(\nu)} - \phi_{R}^{(\nu )})}\,.
\eea
where we introduced the inverse temperatures $\beta_\nu$, chemical potentials $\mu_\nu$, particle number operators $N_R^{(\nu)}$, and thermodynamic grand-potentials $\phi_{R}^{(\nu)} = -\beta^{-1}_\nu \ln \mbox{Tr}\left\{ \exp{e^{-\beta_\nu (H_R^{(\nu)} - \mu_\nu N_{R}^{(\nu)})}} \right\}$ of reservoir $\nu = 1, \dots , N$, $N$ denoting the total number of reservoirs.
This directly (or after suitable transformations) often implies that the Lindblad generator can be additively decomposed in the reservoir index $\nu$.
The master equation in presence of slow driving and multiple reservoirs can thus be formally written as
\bea\label{EQ:lindblad_formal}
\dot{\rho}_S = {\cal L}(t) \rho_S (t) = {\cal L}_0(t) \rho_S (t) + \sum_\nu {\cal L}^{(\nu)}(t) \rho_S (t)\,,
\eea
where ${\cal L}_0(t) \rho_S (t) \hat{=} -\ii \left[H_S(t), \rho_S (t)\right]$ describes the action of the driven system Hamiltonian only.
As discussed before for a single reservoir, the dissipator associated to reservoir $\nu$ will obey detailed balance relations leading to
\bea\label{EQ:local_annihilation}
{\cal L}^{(\nu)}(t) \rho_{\rm eq}^{(\nu)}(t) = 0\,,
\eea
where we have introduced the time-dependent grand-canonical equilibrium state
\bea\label{EQ:gibbs}
\rho_{\rm eq}^{(\nu)}(t) = e^{-\beta_\nu[H_S(t) - \mu_\nu N_S - \phi_{S}^{(\nu )} (t)]}
\eea
in terms of the system Hamiltonian $H_S(t)$, system particle number operator $N_S$, and the system grand-potentials $\phi_{S}^{(\nu)} (t) = -\beta_{\nu}^{-1} \ln \mbox{Tr} \left\{ e^{-\beta_\nu[H_S(t) - \mu_\nu N_S]} \right\}$.

\subsection{Average thermodynamics}\label{average_thermodynamics}

The change of the system energy under the quantum master equation dynamics can be decomposed as
\bea
\dot{E} &=& \frac{d}{dt} \trace{H_S(t) \rho(t)}\nn
&=& \trace{\dot{H}_S \rho} + \sum_\nu \mu_\nu \trace{N_S {\cal L}^{(\nu)} \rho} + \sum_\nu \trace{(H_S(t) - \mu_\nu N_S) {\cal L}^{(\nu)} \rho}\nn
&=& \dot{W} + \sum_\nu \dot{Q}^{(\nu)}\,,
\eea
where we omit the system index $S$ on the density matrix and the time-dependence in the Liouvillians ${\cal L}^{(\nu)}$ for brevity.
The work performed on the system contains a mechanical contribution ($\dot{W}_m$) due to the external driving and a chemical one ($\dot{W}_c$) 
due to the particle transfers with the reservoirs, $\dot W = \dot W_m + \dot W_c$, where
\bea
\dot{W}_m = \trace{\dot{H}_S \rho}\,, \qquad \mbox{and} \qquad \dot W_c = \sum_\nu \mu_\nu \trace{N_S {\cal L}^{(\nu)} \rho}\,.
\eea
The heat current entering the system from reservoir $\nu$ is 
\bea\label{EQ:heat_current}
\dot Q^{(\nu)} = \trace{(H_S(t) - \mu_\nu N_S) {\cal L}^{(\nu)} \rho} \; .
\eea
After having established the first law we now turn to the second law and introduce the von-Neumann entropy which represents the system entropy 
\bea\label{EQ:von_neumann_entropy}
S(t) = - \trace{\rho \ln \rho}.
\eea
Its time evolution is given by
\bea
\dot{S} = - \frac{d}{dt} \trace{\rho \ln \rho} = - \trace{\dot\rho \ln \rho}\,,
\eea
where we used $\trace{\rho \frac{d}{dt} \ln \rho} = 0$ (this can be shown using the fact that the density matrix can be diagonalized by unitary transformation).
Using the Lindblad generator one can directly see that the Hamiltonian driving does not directly contribute to the change of entropy so that we have
\bea\label{EQ:entropy_change}
\dot{S} = -\sum_\nu \trace{\left[{\cal L}^{(\nu)} \rho\right] \ln \rho}\,.
\eea
The entropy production is then given by the sum of the system entropy change plus the entropy change in the reservoirs (caused by the heat flows)
\bea\label{EQ:entropy_production}
\dot{S}_\ii \equiv \dot{S} - \sum_\nu \beta_\nu \dot{Q}^{(\nu)} \, \geq 0.
\eea
This expression can be proven to be positive by using Spohn's inequality~\cite{spohn1978b}, but we also provide a direct proof in Appendix~\ref{entropy_positivity}.

We finish with a note on the Shannon entropy of the system which by construction depends on the basis $S_{\rm Sh} = - \sum_i \rho_{ii} \ln \rho_{ii}$.
For master equations in the rotating wave approximation, the basis chosen is the energy eigenbasis. 
This Shannon entropy does not depend on the eigenstate coherences which anyway evolve independently of the populations. 
Furthermore, it is larger or equal than the von-Neumann entropy.
Indeed, the relative entropy between a density matrix and its diagonal part $\rho_D$ reads
\bea
D(\rho, \rho_D) = \trace{\rho \ln \rho - \rho \ln \rho_D} = -S + S_{\rm Sh} - \trace{(\rho-\rho_D) \ln \rho_D}\,.
\eea
Since $(\rho-\rho_D)$ only contains off-diagonal matrix elements whereas $\ln \rho_D$ has only entries on the diagonal, we have that 
\bea
\trace{(\rho-\rho_D)\ln\rho_D} = \sum_{ij} (\rho-\rho_D)_{ij} \left(\ln\rho_D\right)_{ji}
= \sum_i (\rho-\rho_D)_{ii} \left(\ln\rho_D\right)_{ii} = 0\,.
\eea
Since the relative entropy is non-negative $D(\rho,\rho_D)\ge 0$ under dynamics generated by a Lindblad master equation as we consider here, it follows that $S \le S_{\rm Sh}$.
Note however that $\dot{S}$ and $\dot{S}_{\rm Sh}$ do not obey a general inequality.
Similarly, the correct entropy production rate~(\ref{EQ:entropy_production}) and a Shannon-based entropy production rate 
$\dot{S}_\ii^{\rm Sh} = \dot{S}_{\rm Sh} - \sum_\nu \beta_\nu \dot{Q}^{(\nu)}$ are not generally related by an inequality.

\subsection{Fluctuating thermodynamics}\label{stochastic_thermodynamics}

\subsubsection{Counting statistics}\label{counting_statistics}

Within the same approximations used to derive the quantum master equation, one can derive the full counting statistics for the energy and matter transfers using the dressed master equation formalism \cite{Esposito_2009_ReviewsofModernPhysics, schaller2009b}. 
The measurement scheme corresponds to two point projective measurements of the energy $H^{(\nu)}_R$ and particle number $N^{(\nu)}_R$ in the reservoirs $\nu = 1, \dots , N$. 
The energy and particle transfer generating function $G (\{ \xi_\nu \}, \{ \lambda_\nu \}, t)$ is then obtained by taking the trace of the dressed density matrix of the system $\rho (\{ \xi_\nu \} , \{ \lambda_\nu \} , t)$
\bea
G (\{ \xi_\nu \}, \{ \lambda_\nu \} , t) = \mbox{Tr} \left\{ \rho (\{ \xi_\nu \}, \{ \lambda_\nu \} , t) \right\},
\eea
where the counting field vectors $\{ \xi_\nu \} = \{ \xi_1 , \xi_2 , \dots, \xi_N \}$ and $\{ \lambda_\nu \} = \{ \lambda_1 , \lambda_2 , \dots, \lambda_N \}$ account for, respectively, the energy and matter currents out of the reservoirs. 
The dressed system density matrix satisfies the dressed quantum master equation 
\bea\label{EQ:bms_lindblad_CF}
\dot\rho (\{ \xi_\nu \}, \{ \lambda_\nu \} , t) &=& {\cal L} (\{ \xi_\nu \}, \{ \lambda_\nu \} , t) \rho (\{ \xi_\nu \}, \{ \lambda_\nu \} , t)\\
&& \hspace{-2.5cm} \equiv -\ii \left[H_S (t) + \sum_{ab} \sum_{\tilde \nu} \sigma^{(\tilde \nu)}_{ab} (t) L_{ab} (t), \rho  (\{ \xi_\nu \}, \{ \lambda_\nu \} , t) \right] + \sum_{ab,cd} \sum_{\tilde \nu} \gamma^{(\tilde \nu)}_{ab,cd} (t) \times\nn
 && \hspace{-2cm} \times \left[C_{ab,cd}(\xi_{\tilde \nu},\lambda_{\tilde \nu}, t) L_{ab} (t) \rho (\{ \xi_\nu \}, \{ \lambda_\nu \} , t) L_{cd}^\dagger (t) - \frac{1}{2} \left\{L_{cd}^\dagger (t) L_{ab} (t) , \rho (\{ \xi_\nu \}, \{ \lambda_\nu \} , t) \right\}\right]\,,\nonumber
\eea
whose dressed Liouvillian depends on the counting fields. The factors
\bea \label{counting_field_factor}
C_{ab,cd}(\xi_\nu ,\lambda_\nu , t) = \exp\left\{ \left[\ii\xi_\nu (E_b (t) -E_a (t) ) + \ii\lambda_\nu (N_b-N_a)\right]\right\}\,,
\eea
contain the counting fields keeping track of the energy and matter transfers with the reservoirs. 
The dressed quantum master equation (\ref{EQ:bms_lindblad}) reduces to the regular quantum master equation for the system reduced density matrix when the counting fields are set equal to zero, i.e.s $\{ \xi_\nu \} = \{ \lambda_\nu \} = \{0\}$.

The joined distribution for the energy and matter currents out of the reservoirs \\
$P ( \{ J^{(\nu)}_{E} \}, \{ J^{(\nu)}_{M}\} , t)$ is obtained by using the Fourier transform
\begin{multline}
P(\{ J^{(\nu)}_{E} \}, \{ J^{(\nu)}_{M}\}  , t) \\
 =  \int_{-\infty}^{\infty}  \left[ \prod_\nu t \frac{d \xi_\nu }{2 \pi} \right]   \int_{0}^{2 \pi}  \left[ \prod_\nu t \frac{d \lambda_\nu}{2 \pi} \right]  \, e^{\ii \sum_{\nu}\left( \xi_\nu \Delta E_\nu + \lambda_\nu \Delta N_\nu \right)} G(\{ \xi_\nu \}, \{ \lambda_\nu \} , t),
\end{multline}
where $\Delta E_\nu$ and $\Delta N_\nu$ are the energy and particle number changes in reservoir $\nu$ over a duration $t$, and $J^{(\nu)}_E = \Delta E_\nu / t$ 
and $J^{(\nu)}_M = \Delta N_\nu / t$ denote the corresponding energy and matter currents, respectively.

To calculate the counting statistics of the mechanical work, a projective measurement in the system Hamiltonian $H_S (t)$ is required.
The generating function for the associated counting statistics can be written as \cite{Esposito_2009_ReviewsofModernPhysics}
\bea
G(\alpha ,t) = \mbox{Tr} \left\{ \mbox{e}^{\ii \alpha H_S (t)} \left( {\cal T} \exp \int_{0}^{t} d\tau \, {\cal L} (\tau ) \right) \left( \mbox{e}^{-\ii \alpha H_S (0)} \rho (0) \right) \right\}\,,
\eea
where the counting field $\alpha$ counts the energy changes in the system.

Since the mechanical work is the system energy change minus the total energy which has flown to the reservoirs, the generating function for mechanical power and energy and matter currents can be written as \cite{Silaev_2014_PhysicalReviewE}
\begin{multline} \label{full_GF}
G(\alpha , \{ \xi_\nu \}, \{ \lambda_\nu \} ,t) \\
 = \mbox{Tr} \left\{ \mbox{e}^{\ii \alpha H_S (t)} \left( {\cal T} \exp \int_{0}^{t} d\tau \, {\cal L} (\{ \xi_\nu - \alpha \} , \{ \lambda_\nu \} ,\tau ) \right) \left( \mbox{e}^{-\ii \alpha H_S (0)} \rho (0) \right) \right\} ,
\end{multline}
where $\alpha$ is now the mechanical work counting field. Furthermore, ${\cal T} \exp \left\{ \cdot \right\}$ denotes the time-ordered exponential and $\rho (0)$ the initial density matrix of the system. By Fourier transform we get the corresponding probability distribution
\bea
&&P( w, \{ J^{(\nu)}_{E} \}, \{ J^{(\nu)}_{M}\} , t) = \\
&&\hspace{1.5cm} \int_{-\infty}^{\infty} \frac{d \alpha}{2 \pi}  \int_{-\infty}^{\infty}  \left[ \prod_\nu t \frac{d \xi_\nu }{2 \pi} \right]   \int_{0}^{2 \pi}  \left[ \prod_\nu t \frac{d \lambda_\nu}{2 \pi} \right]  \, e^{ \ii \alpha w + \ii \sum_\nu \left( \xi_\nu \Delta E^{(\nu)} + \lambda_\nu \Delta N^{(\nu)} \right)} G(\alpha , \{ \xi_\nu \}, \{ \lambda_\nu \} , t), \nonumber
\eea
where $w$ denotes the mechanical work performed on the system over time $t$.

\subsubsection{Finite-time fluctuation theorem}\label{finiteTimeFT}


We now consider the generating function (\ref{full_GF}) when the system is driven by a time dependent protocol, $H_S (\tau)$ for $\tau \in \left[ 0, t \right]$, and initially at equilibrium with reservoir $\nu = 1$ 
\bea\label{initialcondition}
\rho_{\rm eq}^{(1)}(0) = e^{-\beta_1[H_S(0) - \mu_1 N_S - \phi_{S}^{(1 )} (0)]}.
\eea
We also consider the corresponding backward process where the system is driven by the time-reversed protocol, $\tilde H_S (\tau) = H_S (t-\tau)$ for $\tau \in \left[ 0, t \right]$, and initially at equilibrium with reservoir $\nu = 1$ at the final time of the forward protocol
\bea\label{initialconditiontm}
\rho_{\rm eq}^{(1)}(t) = e^{-\beta_1[H_S(t) - \mu_1 N_S - \phi_{S}^{(1 )} (t)]}.
\eea
Since the Liouvillian depends parametrically on time through the system Hamiltonian, the generating function for the backward process is given by
\begin{multline} \label{time_reversed_generating}
\tilde G( \alpha , \{ \xi_{\nu} \} , \{ \lambda_{\nu} \} , t )  \\
= \mbox{Tr} \left\{ \mbox{e}^{\ii \alpha \tilde H_S (t)} \left( {\cal T} \exp \int_{0}^{t} d\tau \, \tilde {\cal L} ( \{ \xi_{\nu} - \alpha \} , \{ \lambda_{\nu} \} ,\tau ) \right) \left( \mbox{e}^{-\ii \alpha \tilde H_S (0)} \rho_{\rm eq}^{(1)}(t) \right) \right\}\,,
\end{multline}
where $\tilde {\cal L} ( \{ \xi_{\nu} - \alpha \} , \{ \lambda_{\nu} \} ,\tau ) = {\cal L} ( \{ \xi_{\nu} - \alpha \} , \{ \lambda_{\nu} \} , t-\tau )$.
In the following, we take reservoir $\nu = 1$ as a reference for the energy and particle number counting. Accordingly, we set $\xi_1 = \lambda_1 = 0$ and introduce the new counting field vectors $\{ \xi_\nu \}' = \{ \xi_2, \xi_3 , \dots , \xi_N \}$ and $\{ \lambda_\nu \}' = \{ \lambda_2, \lambda_3 , \dots , \lambda_N \}$.
Using the LDB relation~(\ref{EQ:detailed_balance}), we find the symmetry relation
\bea \label{symmetryL}
{\cal L}^{\dagger} (\{ \xi_{\nu} - \alpha \}' , \{ \lambda_{\nu} \}' ,\tau )  = \mbox{e}^{-\beta_1 \mu_1 N_S} {\cal L} ( \{ iA^{\epsilon}_{\nu} + \xi_{\nu} - (\ii \beta_1 + \alpha) \} ', \{ -\ii A^{n}_{\nu} + \lambda_{\nu} \}' ,\tau ) \mbox{e}^{\beta_1 \mu_1 N_S}
\eea
expressed in terms of the thermodynamic affinities 
\bea
A^{E}_{\nu}= \beta_1 - \beta_{\nu}, \qquad A^{N}_{\nu} = -\beta_1 \mu_1 + \beta_{\nu} \mu_{\nu},
\eea
and where ${\cal L}^{\dagger}$ denotes the conjugate transpose in the system Liouville space, that is, $\mbox{Tr} \left\{ |a_1\rangle \langle a_2 | {\cal L}^{\dagger} \left( |a_3\rangle\langle a_4 | \right) \right\}= \mbox{Tr} \left\{ |a_3\rangle \langle a_4 | {\cal L} \left( |a_1\rangle\langle a_2 | \right) \right\}$, where $|a_i \rangle$ for $i=1 , \dots , 4 $ are arbitrary quantum states in the system Hilbert space.

This symmetry (\ref{symmetryL}) combined with the initial conditions~(\ref{initialcondition}) and~(\ref{initialconditiontm}) implies the finite-time fluctuation theorem
\bea\label{ft_proof}
G( \alpha , \{ \xi_{\nu} \}' , \{ \lambda_{\nu} \}' ,t) &=& \mbox{Tr}\left\{ \mbox{e}^{\ii \alpha H_S (t)} 
\left(  {\cal T} \mbox{e}^{\int_{0}^t \tilde{{\cal L}} (\{ \xi_{\nu} - \alpha \}' , \{ \lambda_{\nu} \}' , t-\tau) d\tau } \right) \mbox{e}^{-\ii \alpha H_S (0)} \rho_{\rm eq}^{(1)}(0) \right\}\nn
&=& \mbox{Tr}\left\{ \mbox{e}^{\ii \alpha H_S (0)} \rho_{\rm eq}^{(1)}(0) 
\left(  {\cal T} \mbox{e}^{\int_{0}^t \tilde{{\cal L}}^{\dagger} ( \{ \xi_{\nu} - \alpha \}', \{ \lambda_{\nu} \}' , \tau) d\tau } \right) \mbox{e}^{-\ii \alpha H_S (t)}  \right\}^{*}\nn
&=&\mbox{Tr}\left\{ \mbox{e}^{\ii (\alpha +\ii \beta_1) H_S (0)} \left(  {\cal T} \mbox{e}^{\int_{0}^t {\cal L}^{\dagger} ( \{ \ii A_{\nu}^{E} + \xi_{\nu} - (\ii\beta_1 +\alpha) \}' , \{ -\ii A_{\nu}^{N} + \lambda_{\nu} \}' , \tau) d\tau } \right) \right.\times\nn
&& \times \left. \mbox{e}^{-\ii (\alpha+ \ii \beta_1) H_S (t)}  \rho_{\rm eq}^{(1)}(t) \right\}^{*} \mbox{e}^{-\beta_1 \Delta \phi_1}\nn
&=&  \tilde G( \ii\beta_1 + \alpha , \{ -\ii A^{E}_{\nu} + \xi_{\nu} \}' , \{ -\ii A^{N}_{\nu} + \lambda_{\nu} \}' ,t)^{*} \, \mbox{e}^{-\beta_1 \Delta \phi_1}\nn
&=&  \tilde G( \ii\beta_1 - \alpha , \{ -\ii A^{E}_{\nu} - \xi_{\nu} \}' , \{ -\ii A^{N}_{\nu} - \lambda_{\nu} \}' ,t) \, \mbox{e}^{-\beta_1 \Delta \phi_1} ,
\eea
where $\Delta \phi_{S}^{(1)} = \phi_{S}^{(1)} (t) - \phi_{S}^{(1)} (0)$. 
At the probability level, the finite time fluctuation theorem is given by
\bea \label{fluct_th_distr}
\ln \frac{P(+w , \{+J^{(\nu)}_E \}' ,\{+J^{(\nu)}_M \}' , t)}{\tilde P( -  w , \{ - J^{(\nu)}_E \}' ,  \{- J^{(\nu)}_M \}' , t ) } = \beta_1 (  w - \Delta \phi_{S}^{(1)}) + \sum_{\nu = 2}^{N} \left( A_{\nu}^{E}  J_{E}^{(\nu)} + A_{\nu}^{N} J_{M}^{(\nu)} \right) t\,,
\eea
where $P(+w , \{+J^{(\nu)}_E \}' , \{+J^{(\nu)}_M \}' , t) = \int d J^{(1)}_E \int d J^{(1)}_N P( w , \{ J^{(\nu)}_E \} , \{ J^{(\nu)}_M \} , t)$, 
and \\
$\tilde P( -  w , \{ -J^{(\nu)}_E\}' ,  \{ -J^{(\nu)}_M\}' , t)$ denotes the corresponding probability distribution along the backward process. 
This fluctuation theorem (\ref{fluct_th_distr}) holds for any given time $t$, and is exclusively expressed in terms of the mechanical power and the energy and matter currents. It is the quantum analogue of the classical result derived in Ref.~\cite{cuetara2014exact}.

\section{Degenerate single quantum dot circuit} \label{a_quantum_master_equation_model}

\begin{figure}[ht]
\begin{tabular}{lr}
\includegraphics[width=0.45 \textwidth]{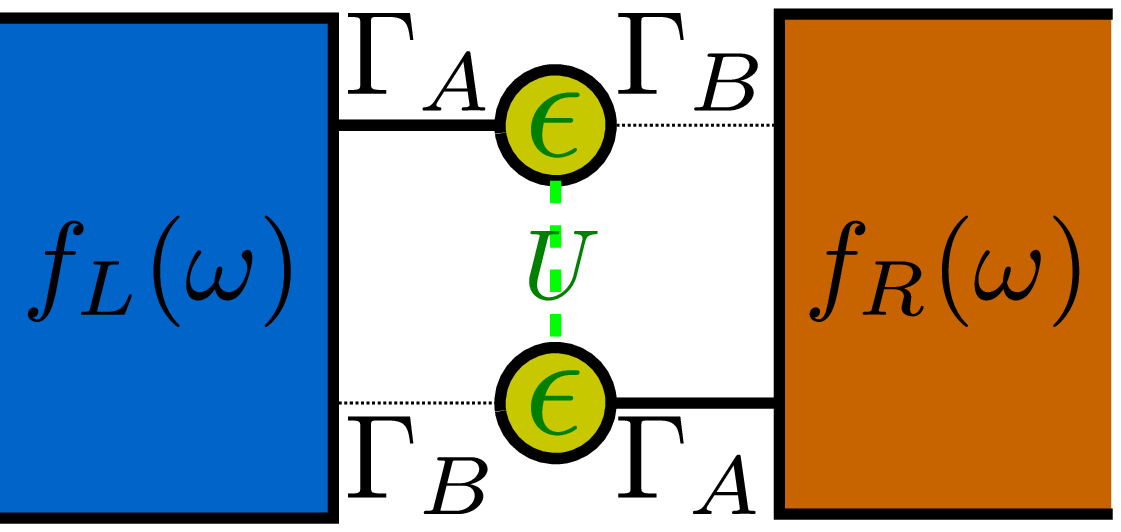}&
\includegraphics[width=0.45 \textwidth]{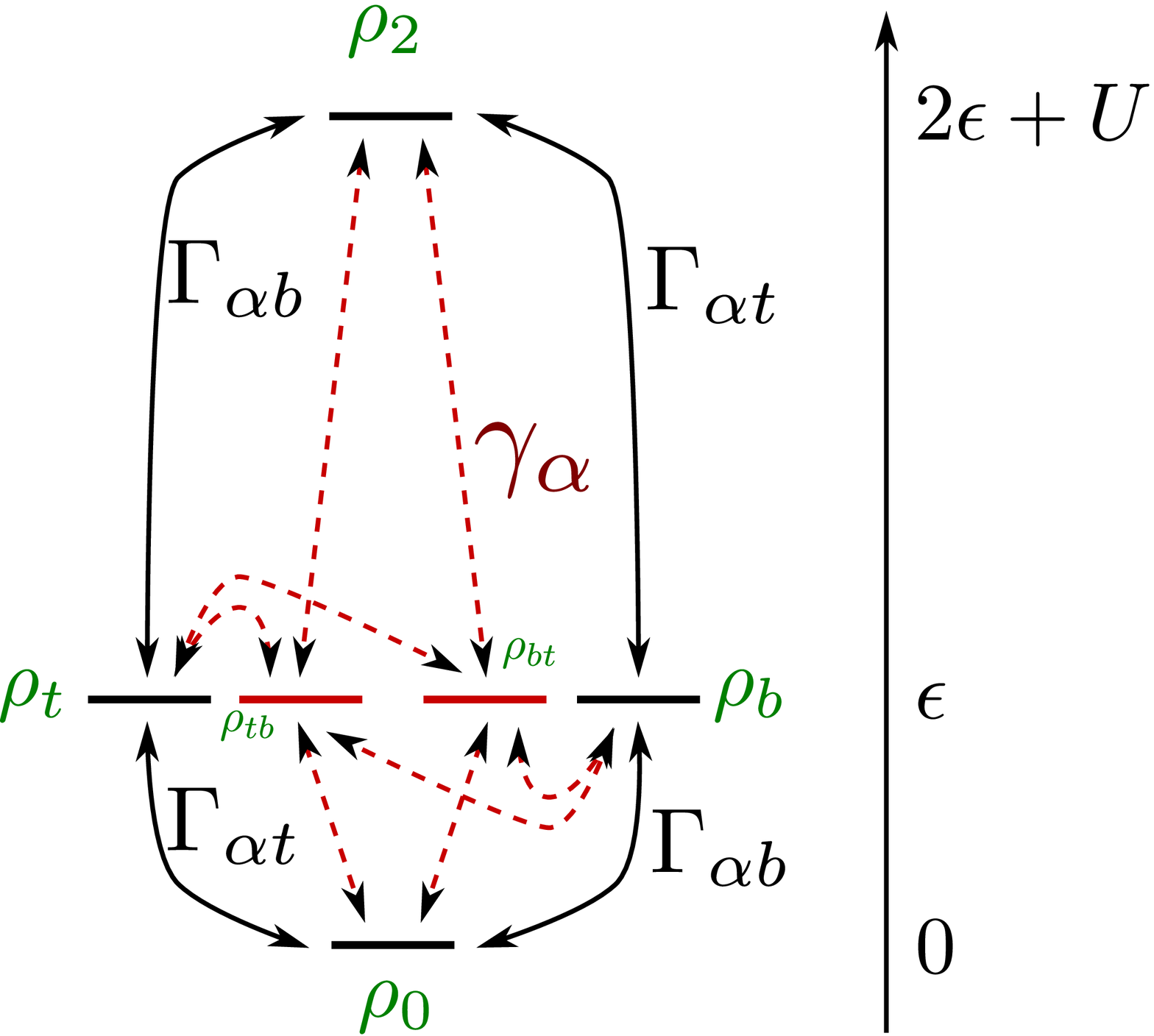}
\end{tabular}
\caption{\label{FIG:masterrategraph}(Color Online)
{\bf Left}: Illustration of the double quantum dot system with degenerate on-site energies $\epsilon$ and Coulomb-interaction $U$ (dashed).
The leads are described by Fermi functions $f_{L/R}(\omega)$ that depend on lead temperatures and chemical potentials.
The peculiar feature of the system is that it is possible to tunnel directly into a superposition of the singly-charged states, 
described by the rate $\gamma=\sqrt{\Gamma_A \Gamma_B}$.
Mainly for simplicity, we consider in this paper a tunnel-coupling configuration with only two different tunneling rates (bold solid and thin dotted).
To avoid a bistable regime we note that we require $\Gamma_A \neq \Gamma_B$.
{\bf Right}: Graph associated to the master equation~(\ref{EQ:master_explicit}).
Solid arrows correspond to conventional transition rates obeying a LDB
relation for each reservoir $\nu \in \{L, R\}$, they are proportional to $\Gamma_{\nu t/ \nu b}$ as indicated.
Dashed arrows connect populations with the coherences, they do not correspond to traditional rates but vanish as $\gamma_\nu \to 0$, 
thus effectively decoupling populations and coherences in the local basis.
}
\end{figure}

We now illustrate our formalism by considering a specific model consisting of two degenerate quantum dots connected to two electron leads, see Fig. \ref{FIG:masterrategraph}.
After defining the model, we first study its average thermodynamics. 
We then compare its counting statistics with and without eigenstate coherences and show that both satisfy the finite time fluctuation theorem derived above. 

\subsection{Model} \label{model_dyn}

We consider a double quantum dot with no direct tunneling between the dots but exactly degenerate on-site energies.
In general, it is well-known that exact degeneracies may give rise to rich dynamics~\cite{braig2005a,darau2009a}.
For our particular model, it is from a transport perspective also well known that negative differential
conductance may arise from the Coulomb interaction due to coherences~\cite{braun2004a,schultz2009a,schaller2009b,schultz2010a}.
The effect has been observed experimentally~\cite{nilsson2010a} and is also present beyond the sequential 
tunneling regime~\cite{karlstroem2011a}.
A distinctive feature of this system is that the attached fermionic contacts allow for electron jumps into
superposition states.
The system, interaction, and reservoir Hamiltonians read
\bea \label{model_ham}
H_S &=& \epsilon\left(d_t^\dagger d_t + d_b^\dagger d_b\right) + U d_t^\dagger d_t d_b^\dagger d_b\,,\nn
H_I &=& \sum_{\nu\in\{L,R\}} \sum_{i\in\{t,b\}} \sum_k \left[t_{k \nu i} d_i c_{k \nu}^\dagger + t_{k \nu i}^* c_{k \nu} d_i^\dagger\right]\,,\nn
H_R &=& \sum_{k\nu} \epsilon_{k\nu} c_{k\nu}^\dagger c_{k\nu}\,.
\eea
Here, the on-site energies $\epsilon$ of the top ($t$) and bottom ($b$) dot are degenerate and $U$ denotes their Coulomb interaction.
The $t_{k\nu ,i}$ denote the tunneling amplitudes into mode $k$ of lead $\nu$ with energy $\epsilon_{k\nu}$ from dot $i$ (top or bottom).
It is visible that both leads may trigger electronic jumps into both dots.

First, we remark that for charged states, not all superposition states are allowed.
In particular, we cannot form superpositions of differently charged states, such that coherences between e.g., the empty and doubly occupied states can be neglected from the beginning.
Formally, they will evolve in a decoupled (and damped) fashion, but in reality they cannot be created in a system-local state and will therefore vanish throughout.
Denoting the diagonal matrix elements of the empty, the top occupied, the bottom occupied, and the doubly occupied state by
$\rho_0$, $\rho_t$, $\rho_b$, and $\rho_2$, respectively, and the admissible coherences between the singly-charged states by $\rho_{tb}$ and 
$\rho_{bt}=\rho_{tb}^*$, the BMS Lindblad master equation~(\ref{EQ:bms_lindblad}) becomes (see Appendix~\ref{APP:derivation} for more details on the derivation)
\bea\label{EQ:master_explicit}
\dot \rho_0 &=& - \left[(\Gamma_{Lt}+\Gamma_{Lb}) f_L + (\Gamma_{Rt}+\Gamma_{Rb}) f_R\right] \rho_0\nn
&&+ \left[\Gamma_{Lt}(1-f_L)+\Gamma_{Rt}(1-f_R)\right] \rho_t 
+ \left[\Gamma_{Lb}(1-f_L)+\Gamma_{Rb}(1-f_R)\right] \rho_b\nn*
&&+\left[\gamma_L (1-f_L)+\gamma_R (1-f_R)\right] \rho_{tb} 
+\left[\gamma_L^* (1-f_L)+\gamma_R^* (1-f_R)\right] \rho_{bt}\,,\nn
\dot \rho_t &=& -\left[\Gamma_{Lt} (1-f_L) + \Gamma_{Rt} (1-f_R) + \Gamma_{Lb} f_L^U + \Gamma_{Rb} f_R^U \right] \rho_t\nn*
&&+\left[\Gamma_{Lt}f_L + \Gamma_{Rt} f_R\right] \rho_0 + \left[\Gamma_{Lb} (1-f_L^U) + \Gamma_{Rb} (1-f_R^U)\right] \rho_2\nn*
&&+\frac{1}{2} \left[\gamma_L(f_L^U-(1-f_L)) + \gamma_R(f_R^U-(1-f_R)) -\ii\gamma_L \Sigma_L - \ii \gamma_R \Sigma_R\right]\rho_{tb}\nn*
&&+\frac{1}{2} \left[\gamma_L^*(f_L^U-(1-f_L)) + \gamma_R^*(f_R^U-(1-f_R)) +\ii\gamma_L^* \Sigma_L + \ii \gamma_R^* \Sigma_R\right]\rho_{bt}\,,\nn
\dot \rho_b &=& -\left[\Gamma_{Lb}(1-f_L) + \Gamma_{Rb}(1-f_R) + \Gamma_{Lt} f_L^U + \Gamma_{Rt} f_R^U\right]\rho_b\nn*
&&+\left[\Gamma_{Lb}f_L+\Gamma_{Rb} f_R\right] \rho_0 + \left[\Gamma_{Lt}(1-f_L^U) + \Gamma_{Rt}(1-f_R^U)\right] \rho_2\nn*
&&+\frac{1}{2} \left[\gamma_L(f_L^U-(1-f_L)) + \gamma_R(f_R^U-(1-f_R)) + \ii \gamma_L \Sigma_L + \ii \gamma_R \Sigma_R\right]\rho_{tb}\nn*
&&+\frac{1}{2} \left[\gamma_L^*(f_L^U-(1-f_L)) + \gamma_R^*(f_R^U-(1-f_R)) - \ii \gamma_L^* \Sigma_L - \ii \gamma_R^* \Sigma_R\right]\rho_{bt}\,,\nn
\dot \rho_2 &=& -\left[(\Gamma_{Lt}+\Gamma_{Lb})(1-f_L^U) + (\Gamma_{Rt}+\Gamma_{Rb})(1-f_R^U)\right]\rho_2\nn*
&&+\left[\Gamma_{Lb} f_L^U + \Gamma_{Rb} f_R^U\right]\rho_t
+\left[\Gamma_{Lt} f_L^U + \Gamma_{Rt} f_R^U\right]\rho_b\nn*
&&-\left[\gamma_L f_L^U + \gamma_R f_R^U\right]\rho_{tb}
-\left[\gamma_L^* f_L^U + \gamma_R^* f_R^U\right]\rho_{bt}\,,\nn
\dot \rho_{tb} &=& -\frac{1}{2} \Big[(\Gamma_{Lt}+\Gamma_{Lb})(f_L^U+(1-f_L))+(\Gamma_{Rt}+\Gamma_{Rb})(f_R^U + (1-f_R))\nn*
&&\qquad -\ii \Gamma_{Lt} \Sigma_L +\ii \Gamma_{Lb} \Sigma_L-\ii \Gamma_{Rt} \Sigma_R +\ii \Gamma_{Rb} \Sigma_R\Big]\rho_{tb}\nn*
&&+\left[\gamma_L^* f_L + \gamma_R^* f_R\right]\rho_0 - \left[\gamma_L^* (1-f_L^U) + \gamma_R^* (1-f_R^U)\right] \rho_2\nn*
&&+\frac{1}{2} \left[\gamma_L^* (f_L^U - (1-f_L)) + \gamma_R^* (f_R^U - (1-f_R)) - \ii \gamma_L^* \Sigma_L -\ii \gamma_R^* \Sigma_R\right]\rho_t\nn*
&&+\frac{1}{2} \left[\gamma_L^* (f_L^U - (1-f_L)) + \gamma_R^* (f_R^U - (1-f_R)) + \ii \gamma_L^* \Sigma_L +\ii \gamma_R^* \Sigma_R\right]\rho_b\,,\nn
\dot \rho_{bt} &=& \dot{\rho}_{tb}^*\,,
\eea
where we used the wide-band limit for the tunneling rates
\bea\label{EQ:tunnelrates}
\Gamma_{\nu i}(\omega) &=& 2\pi \sum_k \abs{t_{k\nu, i}}^2 \delta(\omega-\epsilon_{k\nu})\to \Gamma_{\nu i}\,,\nn
\gamma_\nu(\omega) &=& 2\pi \sum_k t_{k\nu, t} t_{k\nu, b}^* \delta(\omega-\epsilon_{k\nu}) \to \gamma_\nu\,.
\eea
Whereas the tunneling rates $\Gamma_{\nu i}$ are rates in the traditional sense $\Gamma_{\nu i} \ge 0$ and
describe tunneling processes into top- and bottom-localized electronic states, respectively, this is different for the 
unconventional complex-valued rates $\gamma_\nu$.
Formally, we see that the $\gamma_\nu$ mediate the coupling between coherences and populations and thus allow the system to
jump e.g., from the empty state into a superposition of the singly-charged states.
Depending on the microscopic details of the coupling, the phases of the tunneling amplitudes in Eq.~(\ref{EQ:tunnelrates}) may
interfere destructively (such that $\gamma_\nu \to 0$, which is equivalent to taking the RWA limit) or constructively (when all 
tunneling amplitudes are equal we have $\abs{\gamma_\nu}^2 = \Gamma_{\nu t} \Gamma_{\nu b}$).
This last limit limit of constructive interference $\gamma_\nu\to\sqrt{\Gamma_{\nu t}\Gamma_{\nu b}}$ will be used here as the wide band limit of the secular approximation. 

The thermal reservoir properties are contained in the Fermi functions and Lamb-shift terms
\bea\label{EQ:sigmadef}
f_\nu &=& \frac{1}{e^{\beta_\nu(\epsilon-\mu_\nu)}+1}\,,\qquad
f_\nu^U = \frac{1}{e^{\beta_\nu(\epsilon+U-\mu_\nu)}+1}\,,\nn
\Sigma_\nu &=& \frac{1}{\pi} \Re \left[\Psi\left(\frac{1}{2}+\ii\frac{\beta_\nu(\epsilon+U-\mu_\nu)}{2\pi}\right)
-\Psi\left(\frac{1}{2}+\ii\frac{\beta_\nu(\epsilon-\mu_\nu)}{2\pi}\right)\right]\,,
\eea
where $\Psi(x)$ denotes the digamma function.

We stress a few things before proceeding.
First, as the master equation is of Lindblad form by construction, the density matrix properties will be preserved.
Second, we see that the dissipator is additive in the reservoirs ${\cal L} = {\cal L}_L + {\cal L}_R$.
Each dissipator annihilates its associated Gibbs state, cf. Eq.~(\ref{EQ:local_annihilation}).
Consequently, at global equilibrium ($\beta_L=\beta_R$ and $\mu_L=\mu_R$), the thermal Gibbs state (with vanishing coherences)
is the stationary state.
Finite coherences in the steady state can however arise in nonequilibrium setups, as will be discussed below.
Finally, we mention that the total Liouvillian becomes bistable when $\Gamma_{Lt}=\Gamma_{Lb}=\Gamma_L$ and
$\Gamma_{Rt}=\Gamma_{Rb}=\Gamma_R$, and we will in the following avoid this situation.
The graph of the master equation is depicted in Fig.~\ref{FIG:masterrategraph} right panel.

\subsection{Model thermodynamics} \label{model_thermodynamics}

In what follows we will consider mainly for simplicity the limit 
$\Gamma_{Lt}=\Gamma_{Rb}=\Gamma_A$ and $\Gamma_{Lb}=\Gamma_{Rt}=\Gamma_B$
and $\gamma = \sqrt{\Gamma_A \Gamma_B}$ (or, for the RWA limit, $\gamma=0$), see left panel of Fig.~\ref{FIG:masterrategraph}.
We note we assume $\Gamma_A \neq \Gamma_B$, so that we will not consider the bistable situation in the present paper~\cite{schaller2009b}.

\begin{figure}[h]
\includegraphics[width=0.9\textwidth,clip=true]{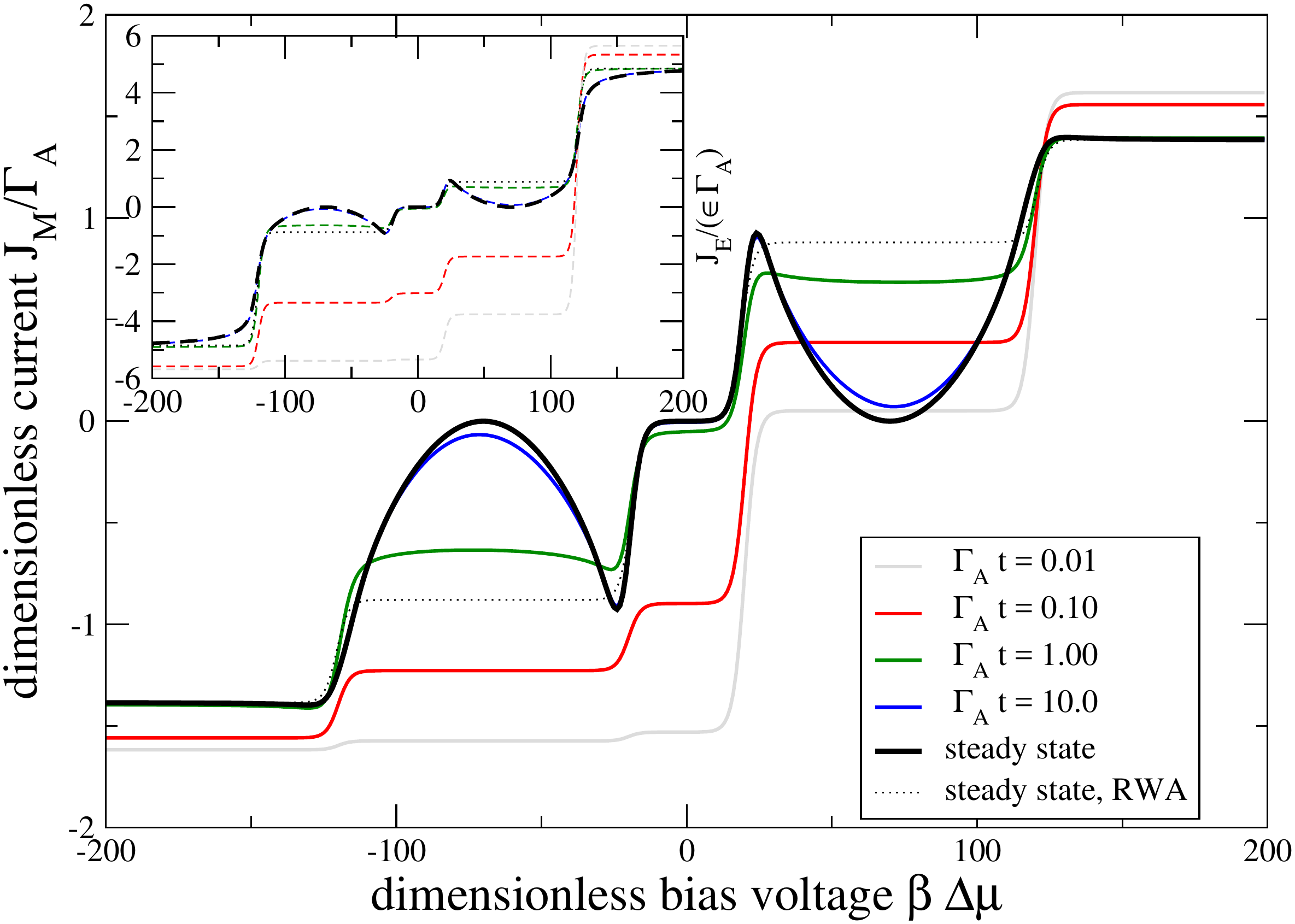}
\caption{\label{FIG:currents}(Color Online)
Plot of the matter current entering the system from the left junction for different times (legend) versus 
dimensionless bias $\Delta \mu = \mu_L-\mu_R$ for the initially completely mixed state.
For small times, the current is not point-symmetric, since the system content dominates the dynamics.
For intermediate times, the current first approaches the steady-state RWA limit (dotted).
For larger times, the coherences induce a large valley of current suppression, with a minimum at $\Delta\mu^*=\pm (2\epsilon+U)$.
The energy currents (inset, same color coding) behave similarly.
Parameters were chosen as $\beta \Gamma_A = 0.1$, $\beta \Gamma_B = 0.225$, $\beta \epsilon = 10$ and $\beta U = 50$.
}
\end{figure}

We can extract the time-dependent energy and matter currents into and from both reservoirs e.g., from Full Counting Statistics methods as discussed in section (\ref{counting_statistics}).
In Appendix~\ref{SEC:liouvillian_cf} we provide the required counting fields exemplarily for transitions triggered by the left junction.
Alternatively, we may also use definitions analogous to the heat current~(\ref{EQ:heat_current}) to calculate energy and matter currents.
In Fig.~\ref{FIG:currents} we plot the time-dependent matter and energy currents for our model versus the potential difference.

Previous investigations of this particular model~\cite{schultz2009a,schaller2009b} have already revealed a 
significant suppression of the steady state matter current due to coherences.
The suppression of the currents is linked to a pure nonequilibrium steady state arising at low temperatures $\beta_L U = \beta_R U \gg 1$ when $\Delta\mu= \mu_L - \mu_R =\pm(2\epsilon+U)$, which
for our particular parameters can be understood analytically, see Appendix~\ref{SEC:current_suppression}.
Here, we complete this picture by the time-dependent evolution and the time-dependent energy current.
Most important, we note the striking difference between the steady-state currents of energy and matter currents of the BMS
(solid and dashed black) and the RWA (dotted black) versions.

\begin{figure}[ht]
\includegraphics[width=0.9\textwidth,clip=true]{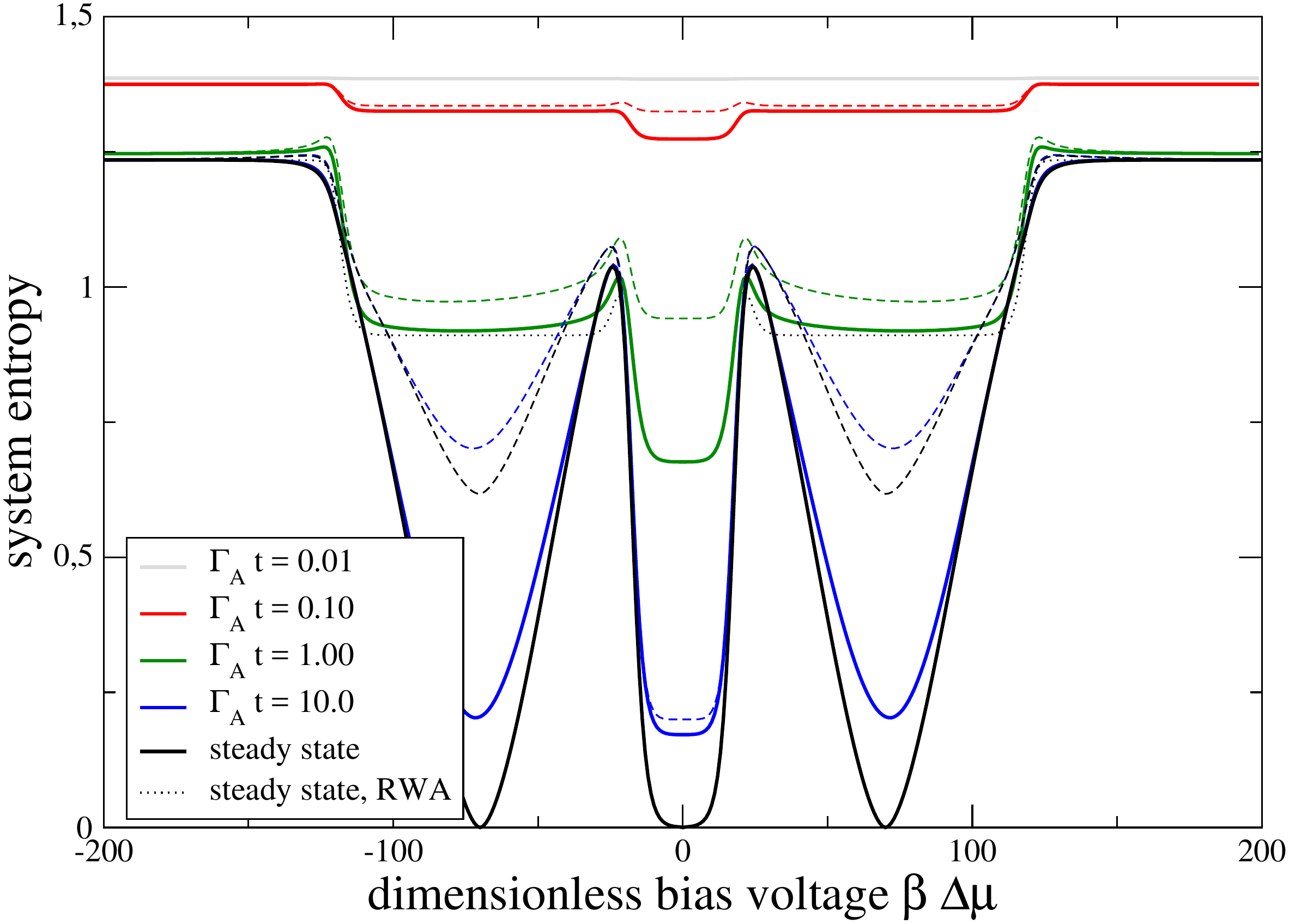}
\caption{\label{FIG:entropy}(Color Online)
Plot of the von-Neumann (solid) and Shannon (dashed) entropies for different times (legend) 
versus dimensionless bias for the initially completely mixed state.
Initially (grey), both entropies are constant and coincide with the maximum value of $\ln(4)$ (dimension of Hilbert space).
As coherences build up, they start to differ until they reach different steady states.
Consistent with the pure delocalized steady state at the current suppression point (see Sec.~\ref{SEC:current_suppression}), the steady-state 
von-Neumann entropy vanishes (solid black) whereas the Shannon entropy does not (dashed black).
The dotted curve shows the steady-state entropy (Shannon) for the RWA rate equation.
Parameters were chosen as in Fig. \ref{FIG:currents}.
}
\end{figure}

Furthermore, from the time-dependent solution of the master equation~(\ref{EQ:bms_lindblad}) we can evidently compute 
the Shannon entropy (in the original energy eigenbasis $\{\ket{0},\ket{t},\ket{b},\ket{2}\}$) and the von-Neumann entropy (basis independent).
Since the first neglects the coherences, these will obviously differ in regions where coherences are present, see Fig.~\ref{FIG:entropy}.
In particular, we can see that the steady-state von-Neumann entropy vanishes when $\Delta\mu=\pm(2\epsilon+U)$, whereas the Shannon
entropy does not, which nicely illustrates that the system reaches a stationary pure state at this nonequilibrium configuration, cf. Appendix~\ref{SEC:current_suppression}.

\begin{figure}[ht]
\includegraphics[width=0.9\textwidth,clip=true]{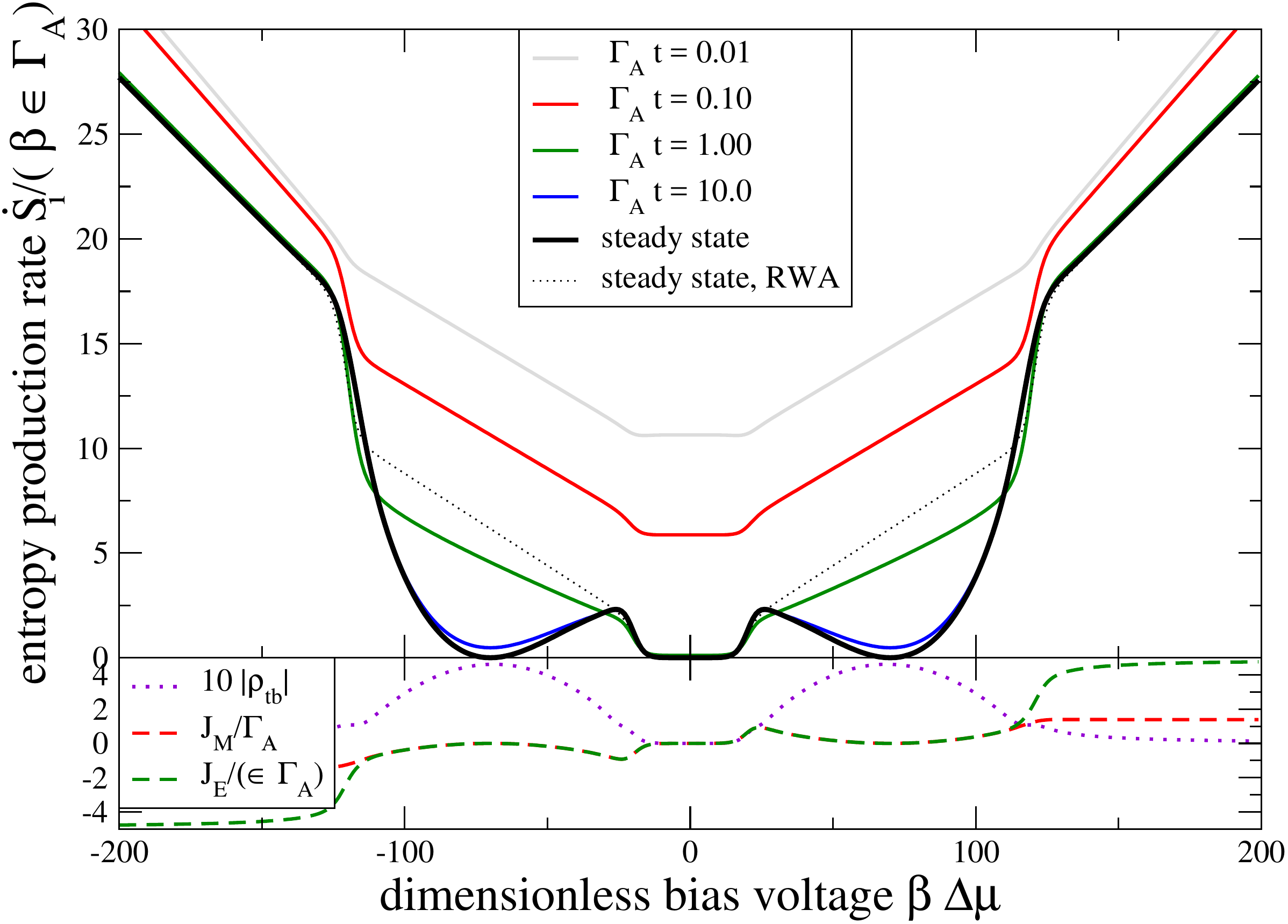}
\caption{\label{FIG:entprodrate}(Color Online)
{\bf Top:} Plot of the (positive) dimensionless entropy production rate for different times (solid curves) versus dimensionless 
bias $\beta \Delta \mu$ for the initially completely mixed state.
For small times, the entropy production rate does not vanish anywhere since the system is not equilibrated.
For large times, the steady-state entropy production rate (bold) is approached, which inherits the minima from
the energy and matter currents (bottom).
In contrast, the RWA version (dotted black) does not exhibit the coherence-induced dips.
{\bf Bottom:} For orientation, we also plot the dimensionless matter (red) and energy (green) currents and the rescaled absolute value
of the coherences (thin dotted magenta).
Parameters were chosen as in Fig. \ref{FIG:currents}.
}
\end{figure}

From the difference between the change of the system entropy and the heat currents we can obtain
the entropy production rate, Eq.~(\ref{EQ:entropy_production}), which we plot in Fig.~\ref{FIG:entprodrate}.
Beyond the evident sanity check that it is positive, we see that even at steady state, coherences between the degenerate states may survive in 
a nonequilibrium setup, which goes along with a suppression of the steady-state entropy production rate.

\begin{figure}[h]
\includegraphics[width=0.9\textwidth,clip=true]{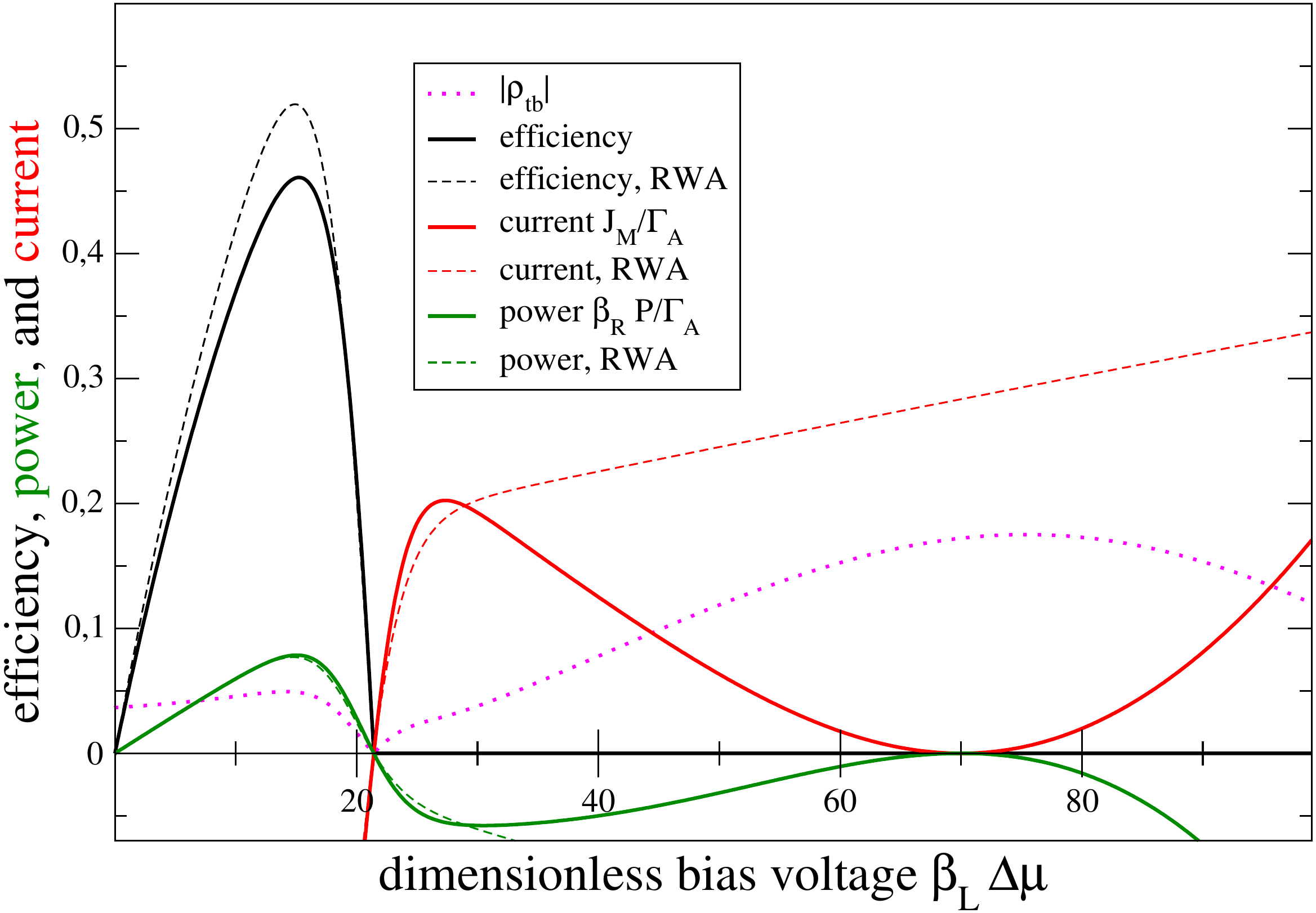}
\caption{\label{FIG:efficiency}
Comparison of efficiency (black), matter current from left to right (red), and generated power (green) for the BMS master equation (solid) and the 
RWA rate equation (thin dashed) versus dimensionless bias voltage.
The thin dotted line denotes the absolute value of the coherence.
The region of finite efficiency is marked by a non-dominating role of coherences in which the RWA and BMS efficiencies are similar.
We notice that the quantum (BMS) efficiency is below the classical (RWA) efficiency.
Parameters were chosen as in Fig. \ref{FIG:currents}.
}
\end{figure}

We now briefly consider how our model can operate as a thermoelectric device. 
We consider the situation in which a thermal gradient is applied between the reservoirs ($\beta_L>\beta_R$) to drive a current against a chemical potential bias ($\Delta \mu = \mu_L - \mu_R >0$).
Denoting the electronic and energy current entering the system from the left reservoir by $J_M$ and $J_E$ (droping the reservoir index $L$), 
the thermoelectric efficiency of this process is defined as the ratio between the generated power $P=-J_M \Delta \mu$ 
and the heat extracted from the hot right reservoir $-(J_E - \mu_R J_M)$ 
\bea
\eta = \frac{J_M \Delta \mu}{J_E - \mu_R J_M}*\Theta(-J_M \Delta \mu )\,.
\eea
The Heaviside function is introduced to indicate that this efficiency is only meaningful in regions of positive power.
Positivity of the steady-state entropy production rate implies -- as usual -- that this efficiency is upper-bounded by the Carnot efficiency, $\eta \leq 1-\beta_R/\beta_L$.
A strong thermoelectric effect requires a large temperature gradient, which in our model reduces the impact of the coherences.
In Fig.~\ref{FIG:efficiency}, we observe numerically that the region of positive power is outside the region where quantum coherences suppress the current.
To obtain a non-negligible power output, we have to consider parameter ranges where the coherences do not significantly modify the energetics. 
Consequently, the BMS and RWA results are qualitatively the same.
In particular the quantum efficiencies ($\gamma_\nu = \sqrt{\Gamma_{\nu t} \Gamma_{\nu b}}$, solid black) and classical efficiencies 
($\gamma_\nu = 0$, dashed black) are rather close, although the quantum efficiency is always smaller than the classical one.
We have numerically observed this inequality also for other parameters.

\subsection{Statistics and fluctuation theorem}

\begin{figure}[ht]
\includegraphics[width=0.9 \textwidth,clip=true]{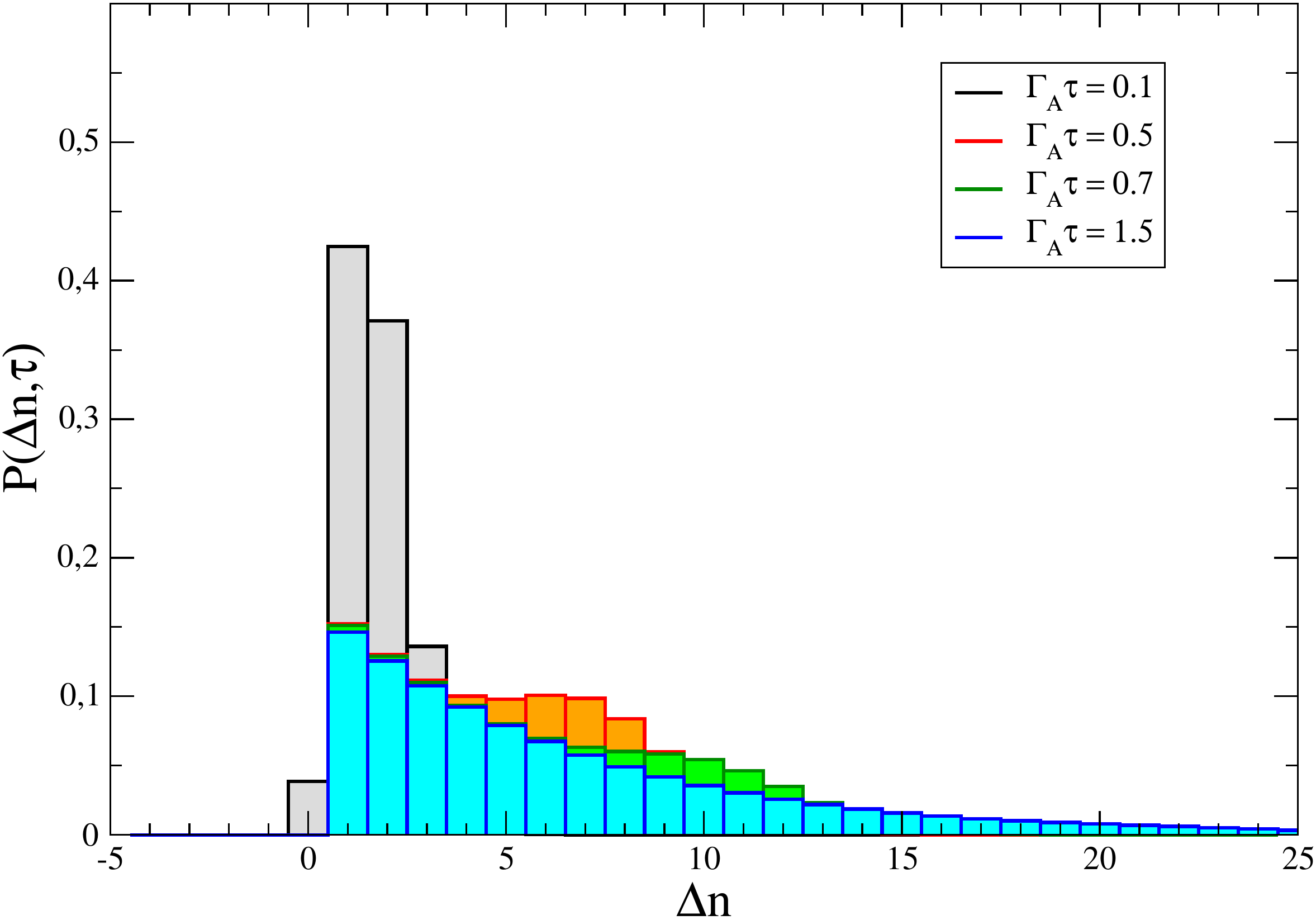}
\caption{\label{FIG:stat_coh}
Probability distributions of the number of particles flowing out of the left reservoir during four different time intervals $\tau$ as obtained from the 
dressed quantum master equation (\ref{EQ:bms_lindblad_CF}) applied to our model. 
The initial condition on the system density matrix is the grand canonical equilibrium distribution with respect to the right 
reservoir (\ref{grand_canonical_eq}). 
The long tail of the long-term distribution (blue) results from telegraph-noise averaging over a $\delta$-peak at $\Delta n=1$ (trapped dark state) 
and a distribution conventionally propagating to the right.
Chemical potentials where chosen as $\beta \mu_L = - \beta \mu_R = 30. $ Other parameters where chosen as in Fig. \ref{FIG:currents}.
}
\end{figure}

\begin{figure}[h]
\includegraphics[width=0.9 \textwidth,clip=true]{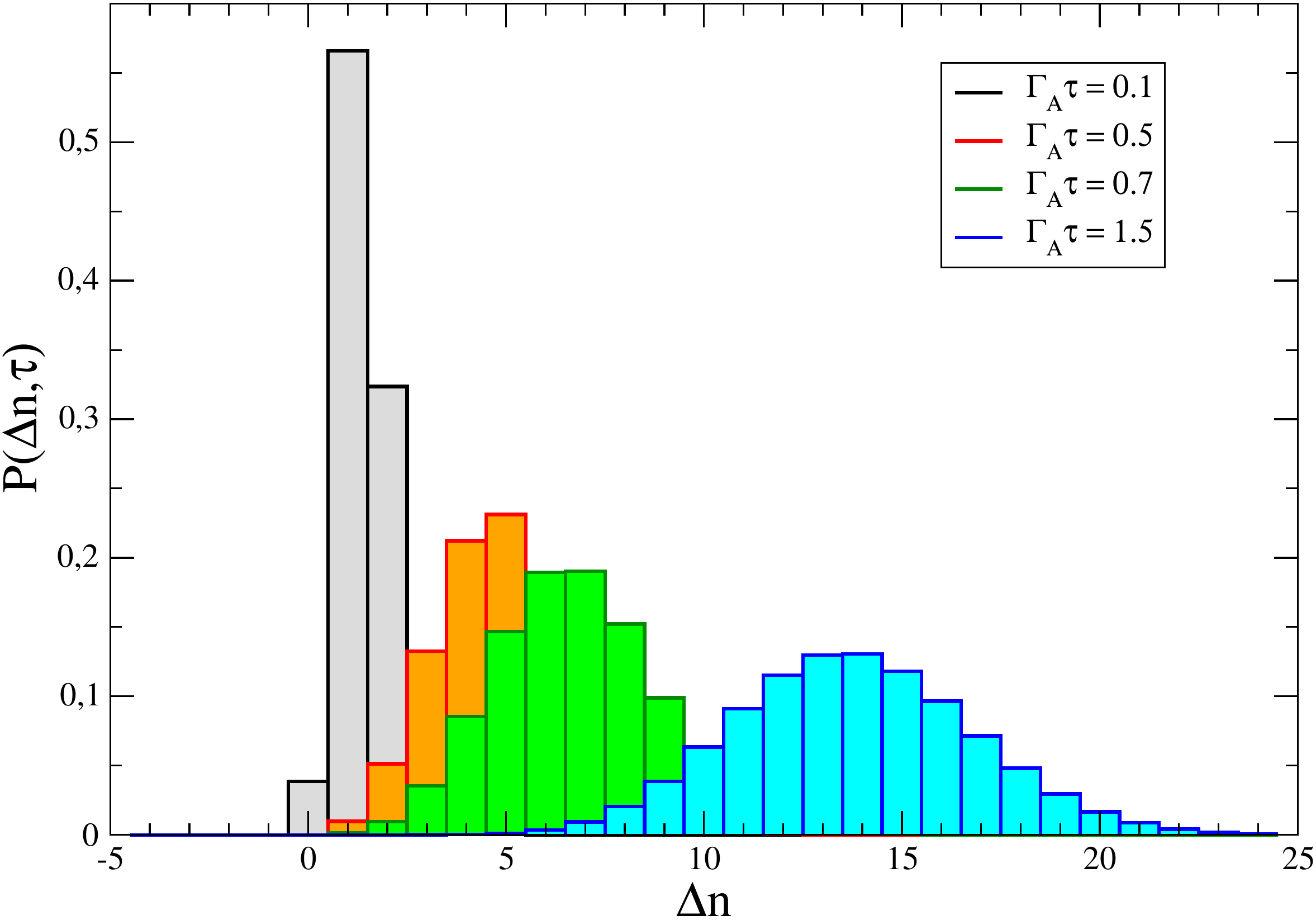}
\caption{\label{FIG:stat_no_coh}
Probability distributions of the number of particles flowing out of the left reservoir during four different time intervals $\tau$ obtained within the rotating wave approximation, that is, neglecting the influence of quantum coherences $\rho_{tb}$ and $\rho_{bt}$ on the statistics. The initial condition on the system density matrix is the grand canonical equilibrium distribution with respect to the right reservoir (\ref{grand_canonical_eq}). 
Chemical potentials where chosen as $\beta \mu_L = - \beta \mu_R = 30. $ Other parameters where chosen as in Fig. \ref{FIG:currents}.
}
\end{figure}

The generating function of work and currents (\ref{full_GF}) can be evaluated numerically by solving the dressed quantum master equation (\ref{EQ:bms_lindblad_CF}) for the specific model (\ref{model_ham}), namely Eq. (\ref{dressedQMEforModel}). The corresponding joined probability distribution is then obtained by a Fourier transform.

As an illustration, we now consider the system introduced in section~\ref{model_dyn} in the isothermal regime $\beta = \beta_L = \beta_R$. 
The initial condition of the system is taken as the grand canonical equilibrium with respect to the right reservoir
\bea\label{grand_canonical_eq}
\rho (0) = \exp{ \left[ -\beta (H_S - \mu_R N_S - \phi_{S}^{R}) \right] }\,,
\eea
where the equilibrium grand-potential is $\phi_{S}^{R} = - \beta^{-1} \ln \mbox{Tr} \left\{ \exp{\left[ -\beta (H_S - \mu_R N_S) \right] }  \right\}$, and where $N_S = d^{\dagger}_t d_t + d^{\dagger}_b d_b$ is the particle number operator in the system.

The distribution $P (\Delta n , \tau)$ of the particle changes in the left reservoir $\Delta n = J_M \tau$ during time $\tau$ is numerically evaluated for three different values of the measurement time. 
The results from the master equation in the secular approximation are compared to those obtained from the RWA master equation (\ref{rwa_mast_eq}), in which one neglects the influence of quantum coherences on the dynamics and current statistics. 
In Fig. \ref{FIG:stat_coh}, we see that the distributions obtained in the former case (i.e., BMS) exhibit a bimodal behavior in the transient regime, which approaches a 
long-tail distribution for large times. 
This was observed in a wide range of parameters close to the current suppression point $\Delta\mu^* = \pm (2 \epsilon + U)$. 
Qualitatively, this can be well understood from the fact that the system is close to a bistable configuration, associated with a near-block form of the Liouvillian: Whereas one block supports a
finite steady-state current, the current associated with the other subspace (with a dark state) vanishes, and telegraph-type averaging over the two distributions 
yields the visible long-tail distribution~\cite{jordan2004a,schaller2010b}.
The diagonal initial state~(\ref{grand_canonical_eq}) then also explains why the long-term distribution starts at $\Delta n=1$: Since the dark state is a 
superposition of the two singly-charged states, at least a single jump event is required to create it.
This effect is totally absent in the latter case (i.e., RWA), where the distribution has the usual bell-shape whose drift gives the finite average current at steady state. The BMS drift instead is, as expected, very small close to the current suppression point when coherences are taken into account (see section \ref{average_thermodynamics}). 
The BMS distribution is thus non-trivial and converges to a distribution with a large tail.
This example shows that not only average currents are affected by sustained coherences, but also their statistics. 

For our choice of initial condition~(\ref{grand_canonical_eq}), the statistics of the current flowing out of the left reservoir must satisfy the fluctuation symmetries~(\ref{ft_proof}) and~(\ref{fluct_th_distr}). 
In the present case, the fluctuation relation (\ref{fluct_th_distr}) reduces to the finite-time fluctuation theorem for the net number of particles transferred to the left reservoir ($\Delta n = J_M \tau$)
\begin{equation} \label{finite_time_FT_hT}
 \ln \frac{P(+\Delta n , \tau)}{P(-\Delta n , \tau)} =\beta \Delta \mu \, \Delta n ,
\end{equation}
where $\Delta \mu = \mu_L - \mu_R$. 
Since we first numerically evaluate the current generating function $G( \lambda , \tau) = \tau^{-1} \sum_{\Delta n} P(\Delta n , \tau) \mbox{e}^{-i \lambda \Delta n}$, and due to the highly oscillating integrals involved in obtaining the distribution $P(\Delta n , \tau)$ at large particle number changes $\Delta n$, it is however simpler to test the equivalent fluctuation 
theorem symmetry (\ref{ft_proof}), which here reduces to
\begin{equation}
G (\lambda, \tau) = G( \ii \beta \Delta \mu - \lambda, \tau).
\end{equation}
This symmetry is indeed verified by the generating functions of the distributions shown in Fig. \ref{FIG:stat_coh}. We note that the fluctuation theorem is also satisfied by the generating function obtained within the RWA (Fig. \ref{FIG:stat_no_coh}) even though the two statistics significantly differ. 
The fact that the statistics obtained within the RWA also satisfies a finite time FT directly results from the fact that transition rates of 
the stochastic master equation (\ref{rwa_mast_eq}) satisfy the LDB relation (\ref{EQ:detailed_balance}).

\section{Summary}\label{Conc}

In the present paper, we established the nonequilibrium thermodynamics of open quantum systems exhibiting degeneracies and described by quantum master equations~(\ref{EQ:bms_lindblad}). 
%
%
We established the first and and second law as well as a finite-time fluctuation theorem solely expressed in terms of the mechanical work and the energy and particle counting statistics. 
Using a simple model with two degenerate quantum dots, we showed that eigenbasis coherences at steady state can generate non-trivial counting statistics such as bi-modality and diverging second and higher cumulants.
These findings will help to elucidate the role of coherences in stochastic thermodynamics. 
A remaining open issue is to be able to treat close-to-degenerate eigenstates within the quantum master equation formalism. 
This is particularly important to treat drivings which can induce crossings between the system eigenenergies.  

\acknowledgments{
G.B.C. is supported by the National Research Fund of Luxembourg (AFR Postdoc Grant 7982468). 
M.E. is supported by the National Research Fund of Luxembourg (project FNR/A11/02) as well as by the European Research Council (project 681456).
G.S is supported by the DFG (SCHA 1646/3-1). 
This work also benefited from the COST Action MP1209.
}

%
\appendix
\section{KMS condition with chemical potentials}\label{APP:kms_chempot}

We essentially just use the invariance of the trace under permutations.
In particular, we can write
\bea
C_{\bar\alpha\alpha}(-\tau-\ii\beta) &=& \frac{1}{Z} \traceR{e^{-\ii H_R(\tau+\ii\beta)} B_{\bar\alpha} e^{+\ii H_R (\tau+\ii\beta)} B_\alpha e^{-\beta H_R} e^{+\beta\mu N_R}}\nn
&=& \frac{1}{Z} \traceR{e^{+\ii H_R \tau} B_\alpha e^{-\ii H_R \tau} e^{+\beta\mu N_R} B_{\bar\alpha} e^{-\beta H_R}}\,.
\eea
The complication for $\mu\neq 0$ is that $N_R$ and $B_{\bar\alpha}$ do not commute.
However, when we compute the sum
\bea
S_\alpha(\tau) &\equiv& \sum_{\bar\alpha} e^{+\beta\mu N_S} A_{\bar\alpha} e^{-\beta\mu N_S} C_{\bar\alpha\alpha}(-\tau-\ii\beta)\nn
&=& \frac{1}{Z} \traceR{e^{+\ii H_R \tau} B_\alpha e^{-\ii H_R \tau} e^{+\beta\mu (N_R + N_S)} \left[\sum_{\bar\alpha} A_{\bar\alpha} B_{\bar\alpha}\right] e^{-\beta\mu N_S} e^{-\beta H_R}}\nn
&=& \frac{1}{Z} \traceR{e^{+\ii H_R \tau} B_\alpha e^{-\ii H_R \tau} \left[\sum_{\bar\alpha} A_{\bar\alpha} B_{\bar\alpha}\right] e^{+\beta\mu N_R} e^{-\beta H_R}}\nn
&=& \sum_{\bar\alpha} A_{\bar\alpha} C_{\alpha\bar\alpha}(\tau)\,,
\eea
we see that we can use that the interaction conserves the total particle number, which proves Eq.~(\ref{EQ:kms_chempot}).
Fourier transformation then yields the relation
\bea
\sum_{\bar\alpha} A_{\bar\alpha} \gamma_{\alpha\bar\alpha}(\omega) = \sum_{\bar\alpha} e^{+\beta\mu N_S} A_{\bar\alpha} e^{-\beta\mu N_S} \gamma_{\bar\alpha\alpha}(-\omega) e^{+\beta\omega}\,.
\eea
Inserting this in the fraction of the dampening coefficients we obtain
\bea
\frac{\gamma_{ab,cd}}{\gamma_{dc,ba}} &=& \frac{\sum_{\alpha\bar\alpha} \gamma_{\alpha\bar\alpha}(E_b-E_a) \bra{a} A_{\bar\alpha} \ket{b} \bra{c} A_\alpha^\dagger \ket{d}^*}
{\sum_{\alpha\bar\alpha} \gamma_{\bar\alpha\alpha}(-(E_b-E_a)) \bra{a} A_{\bar\alpha} \ket{b} \bra{c} A_\alpha^\dagger \ket{d}^*}\nn
&=& \frac{\sum_{\alpha\bar\alpha} \gamma_{\alpha\bar\alpha}(E_b-E_a) \bra{a} A_{\bar\alpha} \ket{b} \bra{c} A_\alpha^\dagger \ket{d}^*}
{\sum_\alpha \bra{a} \left[\sum_{\bar\alpha} \gamma_{\bar\alpha\alpha}(-(E_b-E_a)) A_{\bar\alpha} \right]\ket{b} \bra{c} A_\alpha^\dagger \ket{d}^*}\nn
&=& \frac{\sum_{\alpha\bar\alpha} \gamma_{\alpha\bar\alpha}(E_b-E_a) \bra{a} A_{\bar\alpha} \ket{b} \bra{c} A_\alpha^\dagger \ket{d}^*}
{e^{-\beta(E_b-E_a)} \sum_\alpha \bra{a} \left[\sum_{\bar\alpha} \gamma_{\alpha\bar\alpha}(E_b-E_a) e^{-\beta\mu N_S} A_{\bar\alpha} e^{+\beta\mu N_S} \right]\ket{b} \bra{c} A_\alpha^\dagger \ket{d}^*}\nn
&=& e^{\beta(E_b-E_a)} e^{-\beta\mu(N_b-N_a)}\,,
\eea
which proves Eq.~(\ref{EQ:detailed_balance}).

\section{Positivity of entropy production}\label{entropy_positivity}

In order to establish the positivity of the entropy production defined by (\ref{EQ:entropy_production}), we first note that the heat flow (\ref{EQ:heat_current}) out of reservoir $\nu$ can be written as
\bea\label{EQ:entropy_flow2}
\beta_\nu \dot{Q}^{(\nu)} = - \trace{ \left[{\cal L}^{(\nu)} \rho\right] \left[\ln \rho^{(\nu)}_{\rm eq}\right] }.
\eea
The entropy production itself can then be expressed as
\bea\label{EQ:entropy_production2}
\dot{S}_\ii = -\sum_\nu \trace{\left[{\cal L}^{(\nu)} \rho\right] \left[\ln \rho - \ln \rho^{(\nu)}_{\rm eq}\right]}\, \geq 0.
\eea
Spohn's inequality~\cite{spohn1978b} states that each individual $\nu$ contribution in this last expression is non-negative, 
but we demonstrate this explicitly below.

Completely positive and trace-preserving maps -- like the evolution $V$ generated by Lindblad generators -- are contractive, i.e., they decrease the distance between any two states $D(V A, V B) \le D(A, B)$. This also holds for more general distances such as the quantum relative entropy~\cite{lindblad1975a}
\bea
D(\rho \parallel \sigma) \equiv \trace{\rho \left[\ln \rho - \ln \sigma\right]}\,.
\eea
Choosing $A=\rho(t)$, $B=\rho^{(\nu)}_{\rm eq}(t)$, and $V(t+\Delta t, t)$ as the propagator associated to $\dot\rho={\cal L}^{(\nu)}(t) \rho$ from time $t$ to $t+\Delta t$, it follows that $V(t+\Delta t, t) \rho(t) = \rho(t+\Delta t)$ by construction and $V(t+\Delta t, t) \rho^{(\nu)}_{\rm eq}(t) = \rho^{(\nu)}_{\rm eq}(t) + \ord\{\Delta t^2\}$.
Consequently, we have
\bea
0 &\ge& \frac{1}{\Delta t} \left[D(V(t+\Delta t, t) \rho(t) \parallel V(t+\Delta t, t) \rho^{(\nu)}_{\rm eq}(t)) - D(\rho(t) \parallel \rho^{(\nu)}_{\rm eq}(t))\right]\nn
&=& \frac{1}{\Delta t} \left[D(\rho(t+\Delta t) \parallel \rho^{(\nu)}_{\rm eq}(t)+\ord\{\Delta t^2\}) - D(\rho(t) \parallel \rho^{(\nu)}_{\rm eq}(t))\right]\nn
&=& \frac{1}{\Delta t} \Big[\trace{\rho(t+\Delta t) \ln \rho(t+\Delta t)} - \trace{\rho(t+\Delta t) \ln \rho^{(\nu)}_{\rm eq}(t)}\nn
&&- \trace{\rho(t) \ln \rho(t)} + \trace{\rho(t) \ln \rho^{(\nu)}_{\rm eq}(t)}\Big] + \ord\{\Delta t\}\nn
&\stackrel{\Delta t \to 0}{\longrightarrow}& \frac{d}{dt} \trace{\rho \ln \rho} - \trace{\dot{\rho} \ln \rho^{(\nu)}_{\rm eq}}
= \trace{\dot\rho \left[\ln \rho - \ln \rho^{(\nu)}_{\rm eq}\right]}\nn
&=& \trace{\left[{\cal L}^{(\nu)} \rho \right] \left[\ln \rho - \ln \rho^{(\nu)}_{\rm eq}\right]},
\eea
which establishes the positivity of the entropy production rate (\ref{EQ:entropy_production}).

\section{Details for the specific model}\label{APP:derivation}

In usual derivations of master equations one assumes a tensor-product decomposition of the interaction Hamiltonian, implying that system and reservoir
operators commute.
For fermionic transport, this is obviously not the case as the fermionic operators on system and reservoir anti-commute.
However, it can be checked that the fermionic nature of these operators can be implemented with Pauli matrices
$d_t = \sigma^+ \otimes \f{1} \otimes \f{1}_{\rm rest}$, 
$d_b = \sigma^z \otimes \sigma^+ \otimes \f{1}_{\rm rest}$, and $c_{k\nu} = \sigma^z \otimes \sigma^z \otimes \tilde c_{k\nu}$, 
where the fermionic operators $\tilde c_{k\nu}$ now only act on the reservoir Hilbert space.
To restore the fermionic character in the system, we introduce $\tilde d_1 = - \sigma^+ \otimes \sigma^z$ and $\tilde d_2 = -\f{1}\otimes \sigma^+$, such that the Hamiltonians become
\bea
H_S &=& \epsilon\left(\tilde d_t^\dagger \tilde d_t + \tilde d_b^\dagger \tilde d_b\right) + U \tilde d_t^\dagger \tilde d_t \tilde d_b^\dagger \tilde d_b\,,\nn
H_I &=& \sum_{i\in\{t,b\}} \sum_{\nu\in\{L,R\}} \left[\tilde d_i \otimes \sum_{k\nu} t_{k\nu i} \tilde c_{k\nu}^\dagger + \tilde d_i^\dagger \otimes \sum_{k\nu} t_{k\nu i}^* \tilde c_{k\nu}\right]\,,\nn
H_R &=& \sum_{k\nu} \epsilon_{k\nu} \tilde c_{k\nu}^\dagger \tilde c_{k\nu}\,,
\eea
which appears nearly identical, but now with a tensor product decomposition in the interaction Hamiltonian.
In what follows, we will drop the $\tilde{ }$-superscript and perform the mapping tacitly.

\subsection{Reservoir Correlation Functions}

We have of course the freedom to label the coupling operators in any desired order.
For our model, we choose the coupling operators as 
\bea
A_1 &=& d_t ,\qquad B_1 = \sum_k t_{kL,t} c_{kL}^\dagger\,,\qquad
A_2 = d_t^\dagger ,\qquad B_2 = \sum_k t_{kL,t}^* c_{kL}\,,\nn
A_3 &=& d_b ,\qquad B_3 = \sum_k t_{kL,b} c_{kL}^\dagger\,,\qquad
A_4 = d_b^\dagger ,\qquad B_4 = \sum_k t_{kL,b}^* c_{kL}\,,\nn
A_5 &=& d_t ,\qquad B_5 = \sum_k t_{kR,t} c_{kR}^\dagger\,,\qquad
A_6 = d_t^\dagger ,\qquad B_6 = \sum_k t_{kR,t}^* c_{kR}\,,\nn
A_7 &=& d_b ,\qquad B_7 = \sum_k t_{kR,b} c_{kR}^\dagger\,,\qquad
A_8 = d_b^\dagger ,\qquad B_8 = \sum_k t_{kR,b}^* c_{kR}\,.
\eea
From these definitions, we see that of the $64$ possible, only $16$ correlation functions are non-vanishing, which 
can be written (performing the continuum limit) as
\bea
C_{12}(\tau) &=& \frac{1}{2\pi} \int \Gamma_{Lt}(\omega) f_L(\omega) e^{+\ii\omega\tau} d\omega\,,\qquad
C_{21}(\tau) = \frac{1}{2\pi} \int \Gamma_{Lt}(\omega) [1-f_L(\omega)] e^{-\ii\omega\tau} d\omega\,,\nn
C_{34}(\tau) &=& \frac{1}{2\pi} \int \Gamma_{Lb}(\omega) f_L(\omega) e^{+\ii\omega\tau} d\omega\,,\qquad
C_{43}(\tau) = \frac{1}{2\pi} \int \Gamma_{Lb}(\omega) [1-f_L(\omega)] e^{-\ii\omega\tau} d\omega\,,\nn
C_{14}(\tau) &=& \frac{1}{2\pi} \int \gamma_L(\omega) f_L(\omega) e^{+\ii\omega\tau} d\omega\,,\qquad
C_{41}(\tau) = \frac{1}{2\pi} \int \gamma_L(\omega) [1-f_L(\omega)] e^{-\ii\omega\tau} d\omega\,,\nn
C_{32}(\tau) &=& \frac{1}{2\pi} \int \gamma_L^*(\omega) f_L(\omega) e^{+\ii\omega\tau} d\omega\,,\qquad
C_{23}(\tau) = \frac{1}{2\pi} \int \gamma_L^*(\omega) [1-f_L(\omega)] e^{-\ii\omega\tau} d\omega\,,\nn
C_{56}(\tau) &=& \frac{1}{2\pi} \int \Gamma_{Rt}(\omega) f_R(\omega) e^{+\ii\omega\tau} d\omega\,,\qquad
C_{65}(\tau) = \frac{1}{2\pi} \int \Gamma_{Rt}(\omega) [1-f_R(\omega)] e^{-\ii\omega\tau} d\omega\,,\nn
C_{78}(\tau) &=& \frac{1}{2\pi} \int \Gamma_{Rb}(\omega) f_R(\omega) e^{+\ii\omega\tau} d\omega\,,\qquad
C_{87}(\tau) = \frac{1}{2\pi} \int \Gamma_{Rb}(\omega) [1-f_R(\omega)] e^{-\ii\omega\tau} d\omega\,,\nn
C_{58}(\tau) &=& \frac{1}{2\pi} \int \gamma_R(\omega) f_R(\omega) e^{+\ii\omega\tau} d\omega\,,\qquad
C_{85}(\tau) = \frac{1}{2\pi} \int \gamma_R(\omega) [1-f_R(\omega)] e^{-\ii\omega\tau} d\omega\,,\nn
C_{76}(\tau) &=& \frac{1}{2\pi} \int \gamma_R^*(\omega) f_R(\omega) e^{+\ii\omega\tau} d\omega\,,\qquad
C_{67}(\tau) = \frac{1}{2\pi} \int \gamma_R^*(\omega) [1-f_R(\omega)] e^{-\ii\omega\tau} d\omega\,.
\eea
Above, we have introduced the tunnel rates~(\ref{EQ:tunnelrates}), 
and in particular the $\gamma_\nu(\omega)$ lead to the peculiar physics of the model.
We can directly read off the even Fourier transforms of the correlation functions defined by (\ref{EQ:fourier_transforms})
\bea\label{EQ:ftcorrfunc}
\gamma_{12}(\omega) &=& \Gamma_{Lt}(-\omega) f_L(-\omega)\,,\qquad
\gamma_{21}(\omega) = \Gamma_{Lt}(+\omega) [1-f_L(+\omega)]\,,\nn
\gamma_{34}(\omega) &=& \Gamma_{Lb}(-\omega) f_L(-\omega)\,,\qquad
\gamma_{43}(\omega) = \Gamma_{Lb}(+\omega) [1-f_L(+\omega)]\,,\nn
\gamma_{14}(\omega) &=& \gamma_L(-\omega) f_L(-\omega)\,,\qquad
\gamma_{41}(\omega) = \gamma_L(+\omega) [1-f_L(+\omega)]\,,\nn
\gamma_{32}(\omega) &=& \gamma_L^*(-\omega) f_L(-\omega)\,,\qquad
\gamma_{23}(\omega) = \gamma_L^*(+\omega) [1-f_L(+\omega)]\,,\nn
\gamma_{56}(\omega) &=& \Gamma_{Rt}(-\omega) f_R(-\omega)\,,\qquad
\gamma_{65}(\omega) = \Gamma_{Rt}(+\omega) [1-f_R(+\omega)]\,,\nn
\gamma_{78}(\omega) &=& \Gamma_{Rb}(-\omega) f_R(-\omega)\,,\qquad
\gamma_{87}(\omega) = \Gamma_{Rb}(+\omega) [1-f_R(+\omega)]\,,\nn
\gamma_{58}(\omega) &=& \gamma_R(-\omega) f_R(-\omega)\,,\qquad
\gamma_{85}(\omega) = \gamma_R(+\omega) [1-f_R(+\omega)]\,,\nn
\gamma_{76}(\omega) &=& \gamma_R^*(-\omega) f_R(-\omega)\,,\qquad
\gamma_{67}(\omega) = \gamma_R^*(+\omega) [1-f_R(+\omega)]\,.
\eea
The calculation of the odd Fourier transforms is more involved.
Fortunately, they can be obtained from the even ones by a Cauchy principal value integral
\bea
\sigma_{\alpha\beta}(\omega) = \frac{\ii}{\pi} {\cal P} \int \frac{\gamma_{\alpha\beta}(\bar\omega)}{\omega-\bar\omega} d\bar\omega\,.
\eea

To perform it, we assume that the tunneling rates $\Gamma_{\nu i}(\omega)$ and $\gamma_\nu(\omega)$ can be parametrized by Lorentzian functions
\bea\label{EQ:lorentzian}
\Gamma_{\nu i}(\omega) = \Gamma_{\nu i} \frac{\delta^2}{\omega^2+\delta^2}\,,\qquad
\gamma_\nu(\omega) = \gamma_\nu \frac{\delta^2}{\omega^2+\delta^2}\,.
\eea
Since we will let their width $\delta$ later-on go to infinity, they essentially serve as regulators.
All integrals can then be related to the fundamental integral
\bea\label{EQ:fundamental_integral}
I(\omega) &\equiv& \frac{\ii}{\pi} {\cal P} \int \frac{f(\omega')}{\omega+\omega'} \frac{\delta^2}{\omega'^2 + \delta^2} d\omega'\\
&=& \delta \Big[
\frac{e^{\beta\mu}}{\left(e^{\beta\mu}+e^{\ii\beta\delta}\right)\left(\delta-\ii\omega\right)}
-\frac{\delta}{\left(1+e^{-\beta(\mu+\omega)}\right)\left(\delta^2+\omega^2\right)}\nn
&&+\frac{\ii\delta \Psi\left(\frac{1}{2}-\ii\frac{\beta(\omega+\mu)}{2\pi}\right)
+\frac{1}{2}(\omega-\ii\delta)\Psi\left(\frac{1}{2}-\frac{\beta\delta}{2\pi}-\ii\frac{\beta\mu}{2\pi}\right)
-\frac{1}{2}(\omega+\ii\delta)\Psi\left(\frac{1}{2}+\frac{\beta\delta}{2\pi}-\ii\frac{\beta\mu}{2\pi}\right)}
{\pi\left(\omega^2+\delta^2\right)}
\Big]\,,\nonumber
\eea
where $\Psi(x)$ denotes the digamma function.
It is straightforward to show that the two types of integrals are directly related to the fundamental integral above
\bea
I_a(\omega) &=& \frac{\ii}{\pi} {\cal P} \int \frac{f(-\omega')\Gamma(-\omega')}{\omega-\omega'} = \Gamma I(+\omega)\,,\nn
I_b(\omega) &=& \frac{\ii}{\pi} {\cal P} \int \frac{[1-f(+\omega')]\Gamma(+\omega')}{\omega-\omega'} = \Gamma\left[\ii \frac{\omega \delta}{\omega^2+\delta^2} + I(-\omega)\right]\,.
\eea

\subsection{Liouvillian}\label{SEC:liouvillian_cf}

We now write for our model the dressed Lindblad master equation~(\ref{EQ:bms_lindblad_CF}) describing the dressed system density matrix $\rho (\xi , \lambda ,t)$, where the counting fields $\xi$ and $\lambda$ account for, respectively, the currents of energy and particles out of the left reservoir \cite{Esposito_2009_ReviewsofModernPhysics, schaller2009b}.
We use the {\bf local} energy eigenbasis $\ket{0}$ (empty), $\ket{t}$ (top occupied), $\ket{b}$ (bottom occupied), and $\ket{2}$ (doubly occupied) with system energy
eigenvalues $E_{00}=0$, $E_{t}=\epsilon$, $E_{b}=\epsilon$, and $E_{2} = 2\epsilon+U$, respectively.
We label the dressed density matrix populations as $\rho_0 = \bra{0} \rho \ket{0}$, $\rho_t = \bra{t} \rho \ket{t}$, $\rho_b = \bra{b} \rho \ket{b}$, $\rho_2 = \bra{2} \rho \ket{2}$, and the two relevant coherences as $\rho_{tb} = \bra{t} \rho \ket{b}$ and $\rho_{bt} = \bra{b} \rho \ket{t}$.
We get
\bea \label{dressedQMEforModel}
\dot\rho_{0} &=& -\left[\gamma_{12}(-\epsilon) + \gamma_{34}(-\epsilon) + \gamma_{56}(-\epsilon) + \gamma_{78}(-\epsilon)\right] \rho_{0}\nn*
&&+ \left[\gamma_{21}(+\epsilon)e^{-\ii\lambda-\ii\epsilon\xi} + \gamma_{65}(+\epsilon)\right] \rho_{t} 
+ \left[\gamma_{43}(+\epsilon)e^{-\ii\lambda-\ii\epsilon\xi}+\gamma_{87}(+\epsilon)\right] \rho_{b}\nn*
&&+ \left[\gamma_{41}(+\epsilon)e^{-\ii\lambda-\ii\epsilon\xi} + \gamma_{85}(+\epsilon)\right] \rho_{tb} 
+ \left[\gamma_{23}(+\epsilon)e^{-\ii\lambda-\ii\epsilon\xi}+\gamma_{67}(+\epsilon)\right] \rho_{bt}\,,\nn
\dot\rho_{t} &=& -\left[\gamma_{21}(+\epsilon) + \gamma_{34}(-U-\epsilon) + \gamma_{65}(+\epsilon) + \gamma_{78}(-U-\epsilon)\right]\rho_{t}\nn*
&&+ \left[\gamma_{12}(-\epsilon)e^{+\ii\lambda+\ii\epsilon\xi}+\gamma_{56}(-\epsilon)\right] \rho_{0} 
+ \left[\gamma_{43}(+U+\epsilon)e^{-\ii\lambda-\ii(\epsilon+U)\xi}+\gamma_{87}(+U+\epsilon)\right]\rho_{2}\nn*
&&+\frac{1}{2}\Big[+\gamma_{14}(-U-\epsilon)-\gamma_{41}(+\epsilon)+\gamma_{58}(-U-\epsilon)-\gamma_{85}(+\epsilon)\nn*
&&\qquad-\sigma_{14}(-U-\epsilon)+\sigma_{41}(+\epsilon)-\sigma_{58}(-U-\epsilon)+\sigma_{85}(+\epsilon)\Big]\rho_{tb}\nn*
&&+\frac{1}{2}\Big[-\gamma_{23}(+\epsilon) + \gamma_{32}(-U-\epsilon)-\gamma_{67}(+\epsilon)+\gamma_{76}(-U-\epsilon)\nn*
&&\qquad-\sigma_{23}(+\epsilon)+\sigma_{32}(-U-\epsilon)-\sigma_{67}(+\epsilon)+\sigma_{76}(-U-\epsilon)\Big]\rho_{bt}\,,\nn
\dot\rho_{b} &=&-\left[\gamma_{43}(+\epsilon)+\gamma_{87}(+\epsilon)+\gamma_{12}(-U-\epsilon)+\gamma_{56}(-U-\epsilon)\right]\rho_{b}\nn*
&&+\left[\gamma_{34}(-\epsilon)e^{+\ii\lambda+\ii\epsilon\xi}+\gamma_{78}(-\epsilon)\right]\rho_{0}
+\left[\gamma_{21}(U+\epsilon)e^{-\ii\lambda-\ii(\epsilon+U)\xi}+\gamma_{65}(U+\epsilon)\right]\rho_{2}\nn*
&&+\frac{1}{2}\Big[+\gamma_{14}(-U-\epsilon)-\gamma_{41}(+\epsilon)+\gamma_{58}(-U-\epsilon)-\gamma_{85}(+\epsilon)\nn*
&&\qquad+\sigma_{14}(-U-\epsilon)-\sigma_{41}(+\epsilon)+\sigma_{58}(-U-\epsilon)-\sigma_{85}(+\epsilon)\Big]\rho_{tb}\nn*
&&+\frac{1}{2}\Big[-\gamma_{23}(+\epsilon)+\gamma_{32}(-U-\epsilon)-\gamma_{67}(+\epsilon)+\gamma_{76}(-U-\epsilon)\nn*
&&\qquad+\sigma_{23}(+\epsilon)-\sigma_{32}(-U-\epsilon)+\sigma_{67}(+\epsilon)-\sigma_{76}(-U-\epsilon)\Big]\rho_{bt}\,,\nn
\dot\rho_{2} &=&-\left[\gamma_{21}(U+\epsilon )+\gamma_{43}(U+\epsilon)+\gamma_{65}(U+\epsilon)+\gamma_{87}(U+\epsilon)\right]\rho_{2}\nn*
&&+\left[\gamma_{34}(-U-\epsilon)e^{+\ii\lambda+\ii(\epsilon+U)\xi}+\gamma_{78}(-U-\epsilon)\right]\rho_{t}
+\left[\gamma_{12}(-U-\epsilon)e^{+\ii\lambda+\ii(\epsilon+U)\xi}+\gamma_{56}(-U-\epsilon)\right]\rho_{b}\nn*
&&-\left[\gamma_{14}(-U-\epsilon)e^{+\ii\lambda+\ii(\epsilon+U)\xi}+\gamma_{58}(-U-\epsilon)\right]\rho_{tb}
-\left[\gamma_{32}(-U-\epsilon)e^{+\ii\lambda+\ii(\epsilon+U)\xi}+\gamma_{76}(-U-\epsilon)\right]\rho_{bt}\nn
\dot\rho_{tb} &=&-\Big[+\gamma_{12}(-U-\epsilon)+\gamma_{21}(+\epsilon)+\gamma_{34}(-U-\epsilon)+\gamma_{43}(+\epsilon)\nn*
&&\qquad+\gamma_{56}(-U-\epsilon)+\gamma_{65}(+\epsilon)+\gamma_{78}(-U-\epsilon)+\gamma_{87}(+\epsilon)\nn*
&&\qquad-\sigma_{12}(-U-\epsilon)+\sigma_{21}(+\epsilon)+\sigma_{34}(-U-\epsilon)-\sigma_{43}(+\epsilon)\nn*
&&\qquad-\sigma_{56}(-U-\epsilon)+\sigma_{65}(+\epsilon)+\sigma_{78}(-U-\epsilon)-\sigma_{87}(+\epsilon)\Big]\rho_{tb}\nn*
&&+\left[\gamma_{32}(-\epsilon)e^{+\ii\lambda+\ii\epsilon\xi}+\gamma_{76}(-\epsilon)\right]\rho_{0}
-\left[\gamma_{23}(U+\epsilon)e^{-\ii\lambda-\ii(\epsilon+U)\xi}+\gamma_{67}(U+\epsilon)\right]\rho_{2}\nn*
&&+\frac{1}{2}\Big[-\gamma_{23}(+\epsilon)+\gamma_{32}(-U-\epsilon)-\gamma_{67}(+\epsilon)+\gamma_{76}(-U-\epsilon)\nn*
&&\qquad+\sigma_{23}(+\epsilon)-\sigma_{32}(-U-\epsilon)+\sigma_{67}(+\epsilon)-\sigma_{76}(-U-\epsilon)\Big]\rho_{t}\nn*
&&+\frac{1}{2}\Big[-\gamma_{23}(+\epsilon )+\gamma_{32}(-U-\epsilon)-\gamma_{67}(+\epsilon)+\gamma_{76}(-U-\epsilon)\nn*
&&\qquad-\sigma_{23}(+\epsilon)+\sigma_{32}(-U-\epsilon)-\sigma_{67}(+\epsilon)+\sigma_{76}(-U-\epsilon)\Big]\rho_{b}\,,\nn
\dot\rho_{bt} &=&-\Big[+\gamma_{12}(-U-\epsilon)+\gamma_{21}(+\epsilon)+\gamma_{34}(-U-\epsilon)+\gamma_{43}(+\epsilon)\nn*
&&\qquad+\gamma_{56}(-U-\epsilon)+\gamma_{65}(+\epsilon)+\gamma_{78}(-U-\epsilon)+\gamma_{87}(+\epsilon)\nn*
&&\qquad+\sigma_{12}(-U-\epsilon)-\sigma_{21}(+\epsilon)-\sigma_{34}(-U-\epsilon)+\sigma_{43}(+\epsilon)\nn*
&&\qquad+\sigma_{56}(-U-\epsilon)-\sigma_{65}(+\epsilon)-\sigma_{78}(-U-\epsilon)+\sigma_{87}(+\epsilon)\Big]\rho_{bt}\nn*
&&+\left[\gamma_{14}(-\epsilon)e^{+\ii\lambda+\ii\epsilon\xi}+\gamma_{58}(-\epsilon)\right]\rho_{0}
-\left[\gamma_{41}(U+\epsilon)e^{-\ii\lambda-\ii(\epsilon+U)\xi}+\gamma_{85}(U+\epsilon)\right]\rho_{2}\nn*
&&+\frac{1}{2}\Big[\gamma_{14}(-U-\epsilon)-\gamma_{41}(+\epsilon)+\gamma_{58}(-U-\epsilon)-\gamma_{85}(+\epsilon)\nn*
&&\qquad+\sigma_{14}(-U-\epsilon)-\sigma_{41}(+\epsilon)+\sigma_{58}(-U-\epsilon)-\sigma_{85}(+\epsilon)\Big]\rho_{t}\nn*
&&+\frac{1}{2}\Big[\gamma_{14}(-U-\epsilon)-\gamma_{41}(+\epsilon)+\gamma_{58}(-U-\epsilon)-\gamma_{85}(+\epsilon)\nn*
&&\qquad-\sigma_{14}(-U-\epsilon)+\sigma_{41}(+\epsilon)-\sigma_{58}(-U-\epsilon)+\sigma_{85}(+\epsilon)\Big]\rho_{b}\,.
\eea

When setting the counting fields to zero ($\xi = 0$ and $\lambda = 0$) and using the wide-band limit in the correlation functions~(\ref{EQ:ftcorrfunc}), 
these equations reduce to the quantum master equation for the system density matrix (\ref{EQ:master_explicit}), see also the next section.
We stress that by construction -- though not immediately apparent -- the Lindblad form ensures for preservation of density matrix properties.
For example, the derivative of diagonal density matrix entries must be real-valued, which is ensured by relations among the $\sigma_{ij}$.
The Liouvillian clearly decomposes into left ($\gamma_{ij},\sigma_{ij} : i,j \le 4$) and right ($\gamma_{ij},\sigma_{ij} : i,j \ge 5$) reservoir contributions ${\cal L} = {\cal L}_L + {\cal L}_R$.
One observes that the diagonal thermal state
\bea
(\rho_{0}^\nu,\rho_{t}^\nu, \rho_{b}^\nu,\rho_{2}^\nu) &\propto& (1, 
e^{-\beta_\nu(\epsilon-\mu_\nu)},
e^{-\beta_\nu(\epsilon-\mu_\nu)},
e^{-\beta_\nu(2\epsilon+U-2\mu_\nu)})\,,\nn
\rho_{tb} &=& \rho_{bt} = 0\,, \qquad \mbox{for } \nu = L, R
\eea
is an individual stationary state of the corresponding dissipator, that is, ${\cal L}_\nu \rho^\nu = 0$~\cite{schaller2011a} at vanishing counting fields ($\lambda=0$ and $\xi=0$).
This directly results from the KMS relation of the Fermi functions $1-f_\nu(\omega) = f_\nu(\omega) e^{+\beta_\nu(\omega-\mu_\nu)}$.
The steady state of the complete Liouvillian however will in general not be diagonal.

\subsection{Wideband limit}

We now consider the wideband limit $\delta\to\infty$ in the Lorentzian tunnel-rates~(\ref{EQ:lorentzian}), where 
$\Gamma_{\nu i}(\omega) \to \Gamma_{\nu i}$ and $\gamma_\nu \to \sqrt{\Gamma_{\nu t} \Gamma_{\nu b}}$ 
(admitting a phase for the $\gamma_\nu$ did not lead to observable changes in our model).
The even Fourier transforms of the correlation functions then directly simplify to Fermi functions.
The odd Fourier transforms would individually diverge logarithmically.
However, we see that they always enter in a particular combination
\bea
\Delta \sigma &=& \sigma_{odd,even}(-U-\epsilon)-\sigma_{even,odd}(+\epsilon)\nn
&=& \Gamma \left[I(-U-\epsilon)-I(-\epsilon)-\ii \frac{\epsilon \delta}{\epsilon^2+\delta^2}\right]\,,
\eea
compare Eq.~(\ref{EQ:fundamental_integral}).
In the wide-band limit the divergencies of the individual terms cancel, and we can replace
\bea
\Delta \sigma &\to& \Gamma\left[f(\epsilon) - \frac{\ii}{\pi} \Psi\left(\frac{1}{2}+\ii\frac{\beta(\epsilon-\mu)}{2\pi}\right)\right]\nn
&&-\Gamma\left[f(\epsilon+U) - \frac{\ii}{\pi} \Psi\left(\frac{1}{2}+\ii\frac{\beta(\epsilon+U-\mu)}{2\pi}\right)\right]\nn
&=& \ii \frac{\Gamma}{\pi} \Re \left[\Psi\left(\frac{1}{2}+\ii\frac{\beta(\epsilon+U-\mu)}{2\pi}\right)-\Psi\left(\frac{1}{2}+\ii\frac{\beta(\epsilon-\mu)}{2\pi}\right)\right]\,,
\eea
where the real parts always cancel.
This is quite resistant to further simplification. 
The current suppression occurs when $\mu\to\epsilon+U/2$, where $\Delta \sigma$ vanishes.
Comparing with Eq.~(\ref{EQ:sigmadef}), we see that $\Delta\sigma = \ii \Gamma \Sigma_\nu$.
%

\subsection{Current Suppression Point}\label{SEC:current_suppression}

Now, we will explore the limit of equal temperature $\beta=\beta_L=\beta_R$ but different chemical potentials $\mu_L=+\Delta\mu/2$ and $\mu_R=-\Delta\mu/2$.
In addition, we assume that the bias voltage is tuned to $\Delta\mu \to \Delta\mu^*=2\epsilon+U$ and that the temperature is
very low $\beta U \gg 1$.
If the Coulomb interaction is larger than the on-site energy $U \gg \epsilon$, the Fermi functions either approach zero or one
$f_L \to 1$, $f_L^U \to 0$, $f_R \to 0$, and $f_R^U \to 0$.
Furthermore, we have in this limit that $\Sigma_L \to 0$ and $\Sigma_R \to \ln(3)/\pi$.
Mainly to simplify all expressions, we also consider the limit $\Gamma_{Lt}=\Gamma_{Rb}=\Gamma_A$ and $\Gamma_{Lb}=\Gamma_{Rt}=\Gamma_B$.
The Liouvillian then becomes (with $\gamma=\sqrt{\Gamma_A \Gamma_B}$ and $\Gamma=\Gamma_A+\Gamma_B$)
\bea
{\cal L} = \left(\begin{array}{cccccc}
-\Gamma & \Gamma_B & \Gamma_A & 0 & \gamma & \gamma\\
\Gamma_A & -\Gamma_B & 0 & \Gamma & -\frac{\gamma}{2}-\ii\frac{\gamma\ln 3}{2\pi} & -\frac{\gamma}{2}+\ii\frac{\gamma\ln 3}{2\pi}\\
\Gamma_B & 0 & -\Gamma_A & \Gamma & -\frac{\gamma}{2}+\ii\frac{\gamma\ln 3}{2\pi} & -\frac{\gamma}{2}-\ii\frac{\gamma\ln 3}{2\pi}\\
0 & 0 & 0 & -2\Gamma & 0 & 0\\
\gamma & -\frac{\gamma}{2}-\ii\frac{\gamma\ln 3}{2\pi} & -\frac{\gamma}{2}+\ii\frac{\gamma\ln 3}{2\pi} & -2\gamma & -\frac{\Gamma}{2}-\ii \frac{(\Gamma_A-\Gamma_B)\ln 3}{2\pi} & 0\\
\gamma & -\frac{\gamma}{2}+\ii\frac{\gamma\ln 3}{2\pi} & -\frac{\gamma}{2}-\ii\frac{\gamma\ln 3}{2\pi} & -2\gamma & 0 & -\frac{\Gamma}{2}+\ii \frac{(\Gamma_A-\Gamma_B)\ln 3}{2\pi}
\end{array}\right)\,.
\eea
When $\Gamma_A \neq \Gamma_B$, the nonequilibrium stationary state of this Liouvillian is unique
(near-bistability for $\Gamma_A \approx \Gamma_B$ leads to telegraph-like noise~\cite{schaller2009b}).
It is given by the pure state
\bea
\bar\rho \to 
\left[\sqrt{\frac{\Gamma_A}{\Gamma_A+\Gamma_B}} \ket{t} - \sqrt{\frac{\Gamma_B}{\Gamma_A+\Gamma_B}} \ket{b}\right]
\left[\sqrt{\frac{\Gamma_A}{\Gamma_A+\Gamma_B}} \bra{t} - \sqrt{\frac{\Gamma_B}{\Gamma_A+\Gamma_B}} \bra{b}\right]\,,
\eea
and thus depends on the coupling strengths to both reservoirs.

We note that in this limit, energy and matter currents vanish, since transport requires a mixed steady state.

\end{document}